\documentclass[twocolumn,showpacs,floatfix,aps]{revtex4}
\usepackage{amssymb,amsmath}
\usepackage[dvips]{graphics,color}
\usepackage{epsfig}\usepackage{float}
\usepackage[T1]{fontenc}
\def\be{\begin{equation}}
\def\ee{\end{equation}}
\def\ba{\begin{eqnarray}}
\def\ea{\end{eqnarray}}

\newcommand{\CA}{{\cal A}} \newcommand{\CQ}{{\cal Q}}
\newcommand{\CU}{{\cal U}} \newcommand{\CD}{{\cal D}}
\newcommand{\CL}{{\cal L}} \newcommand{\CE}{{\cal E}}

\begin{document}

\title{A thin disk in higher dimensional space-time and dark matter interpretation}

\author{Carlos H. Coimbra-Ara\'ujo}
\email{carlosc@ifi.unicamp.br}
\affiliation{Instituto de F\'\i sica Gleb Wataghin, Universidade Estadual de Campinas, UNICAMP, 13083-970 Campinas, SP, Brazil.}
\author{Patricio S. Letelier}
\email{letelier@ime.unicamp.br}
\affiliation{Departamento de Matem\'atica Aplicada, Instituto de Matem\'atica, Estat\'\i stica e Computa\c c\~ao Cient\'\i fica, Universidade Estadual de Campinas, UNICAMP, 13083-970, Campinas, SP, Brazil.}

\pacs{04.20.Jb, 04.40.-b, 04.50.+h, 11.25.-w, 95.35.+d}

\begin{abstract}
We find a family of exact solutions of Einstein equations
describing the field of a static axisymmetric thin disk living
in six-dimensional space-time. In particular, we study the disks
constructed (by cutting out the central part of the space-time)
from the conventional Schwarzschild and Chazy-Curzon solutions
with simple extensions in the extra 2 dimensions. The disks are
interpreted in terms of two counter-rotating streams of particles
on free circular orbits. Two extra parameters -- the constants
of motion resulting from projections of the particle-velocities'
extra components -- are constrained by the requirement that the
orbits within the disc be stable. The requirement is met just in
cases when the radial profile of the disc orbital speed fits the
rotation curves of many spiral galaxies. It thus turns out that
the effective modification of the gravity law by extra dimensions
could explain the observed flatness of these curves equally well
as the usually given dark matter interpretation. In the second
part of the paper we show that the inclusion of extra dimensions
also leads to better fits of the gravitational lensing data for
galaxy clusters, without changing results obtained in solar system
scales. Finally, we discuss whether the effect of extra dimensions
could also be translated as following from the occurrence of extra
matter ("Kaluza-Klein particles"). A comment on possible properties
of such particles and a chance to detect them at LHC is given in
the appendix.

\end{abstract}

\maketitle

\section{Introduction}
In  the present work we are interested to model a gravitational $4D$ disk living in a universe endowed with extra dimensions. The motivation comes from the various properties one can derive as rotation curves and density and pressure profiles now carried on with a more general geometrical approach for space-time. A first serious accomplishment at this direction is only achieved by investigating solutions of Einstein field equations in axially symmetric configurations in $D$ dimensions. Exact solutions of Einstein field equations in axially symmetric configurations are an important tool to understand the dynamical properties of real systems which can be described approximately by a thin disk. In this sense, axially symmetric solutions have great astrophysical interest, because they can be used to model galaxies and  accretion disks. A long range of disk solutions was derived with or without radial pressure. Solutions for static disks without radial pressure were first studied by Bonnor and Sackfield \cite{bonnor}, and Morgan and Morgan \cite{morgan1}, and with radial pressure by Morgan and Morgan \cite{morgan2}. A large class of static thin disks solutions were obtained by Letelier and Oliveira \cite{oliveira} using the inverse scattering method. Disks with radial tension have been considered in \cite{gonzalez}, and disk models with electric fields \cite{ledvinka} and magnetic fields \cite{letelier}, and both magnetic and electric fields \cite{magele}. Solutions for self-similar static disks were analyzed by Lynden-Bell and Pineault \cite{lynden}, and Lemos \cite{lemos}. Another relevant approaches are the superposition of static disks with black holes or disk solutions with halos \cite{lemoslet}, and see other important astrophysical solutions to mimic AGNs in \cite{vogt}. Recently Vogt and Letelier refined general relativist models of galaxies in a considerable way \cite{vogtgal}. Important discussions about the role of general relativistic disks to explain rotation curves of galaxies can be found, e.g., in \cite{cooperstock} and essential counter-arguments in \cite{vogtlet}. On axisymmetric geometries in $D$ dimensions, generalized Weyl solutions in $D$ dimensions were presented by Emparan and Reall \cite{emparan} and important objects were classified by them, as black rings and Kaluza-Klein bubbles. About disk rotation curves in higher dimensions see for instance \cite{platonic1}. For other important works about solutions of Einstein equations in $D$ dimensions see \cite{maison}.

The treatment proposed at the present work permits to find important imprints of extra dimensions in the density profile of the disk. A first result obtained by introducing extra dimensions shows important modifications in rotation curves profiles. As rotation curves are the major tool for determining the distribution of mass in disk galaxies, and are also important to study kinematics and to infer the evolutionary histories in galactic systems, as also the most basic and classic manner to bring on the presence of dark matter in galaxies, maybe it is possible that a relativistic disk model embedded in a multidimensional universe shed some light on the dark matter problem. Actually, dark matter (DM) is one of the most intringuing problems in astronomy, astrophysics and cosmology. This puzzle has paramount consequences on the deep comprehension of the universe, matter in general and particle physics in particular. Rotation curves of galaxies and gravitational lensing of galactic clusters are the main basic observational tools to understand the presence of DM in universe \cite{oort}. On this behalf, the nature of DM can be investigated by applying a range of gravitational models: by postulating a halo of exotic matter around galaxies (e.g. CDM models \cite{cdm}) or by modifying gravity at large scales (e.g. MOND, TeVeS and STVG \cite{mond}; as an important review about the subject see, e.g., \cite{mannheim}). Another relevant alternative approach is to enhance astrophysical and cosmological models by using extra dimensions \cite{maartens,antoniadis,kkdm,acd}. Here, to understand the role of dynamics in producing a DM behavior in galactic disks, as a first step, we model and investigate a general relativistic disk and also calculate the gravitational lensing regime for a galactic cluster living in a universe endowed with extra dimensions. We adhere to the standard general relativity, just in space-time with 5 spatial dimensions.

An important question, unanswered until now, is why a galactic disk does not follow a gravitational Newtonian or general relativistic  4D profile for rotation curves. The extradimensional argument is a proposed solution in particle physics field: there the cosmological amount of dark matter is used to calculate cross sections of postulated Kaluza-Klein (KK) particles to be tested in future experiments at colliders \cite{kkdm}. Here we follow another direction to show quantitatively how one can calculate rotation curves from GR in a higher dimensional pattern. Therefore, we construct a disk endowed with extra dimensions by solving the vacuum Einstein equations for an extension of the  Weyl's metric. The main purpose of the work is to construct a top-down model, where we are interested in what is coming to pass at large scales assuming a multidimensional scenario. Usually thin disks are solved from Chazy-Curzon or Schwarzschild solutions and as the simplest example, we solve the case for a $6D$ space-time. The system is constructed from such commonly used solutions: they are employed to generate the disk by introducing discontinuities in the first derivatives of the metric functions (image method). Two constants of motion from projection of extradimensional particle velocities are the free parameters of the model. We prevent the {\it ad hoc} adjustment of such parameters with observed rotation curves, preferring to investigate values where the disk becomes stable. The stability is achieved when the disk is Newtonian-like (where such parameters are null) or for a tiny range of values that astonishingly makes the circular geodesics fit with great precision the rotation curves of many  spiral galaxies. We compare such results to well succeeded astrophysical dark matter profiles and demonstrate that our predictions recover naturally gravity at solar system and give the same gravitational lensing as does a dynamically successful dark halo model. We propose, following such results, that the geometry of a space-time endowed with extra dimensions modify gravity at certain distance/mass scales and dark matter is replaced by the picture of a multidimensional universe. We also consider, for completeness, another interpretation where the geometry can induce Kaluza-Klein fields and particles on the $4D$ space-time, and the present model could be a interesting constraint for theories on KK dark matter particles.

The present work is organized as follows: in Section \ref{sec:4ddisks} we present the principal aspects of static thin disk solutions in $4D$ and the expression for the energy-momentum tensor of the disk. In Section \ref{sec:weylD}, we derive general Weyl's solutions. To derive Weyl's solutions to higher dimensions is to find all solutions of the vacuum Einstein equations that admit $D-2$ orthogonal commuting Killing vector fields. We show, following \cite{emparan}, that in $D$ dimensions, Einstein equations in a Weyl geometry is reduced to the Laplace's equations, what permits to use the image method to construct a disk living in a $D$ dimensional universe. In Section \ref{sec:6D} we exemplify by constructing a simple multidimensional disk model using six dimensional axially symmetric Einstein equations in vacuum, with a  ``Schwarzschild solution'' for the four dimensional part; the extra dimensional part is solved with a Chazy-Curzon solution.  These solutions are employed to generate a disk by introducing discontinuities in the first derivatives of the metric functions (image method). In Section \ref{sec:rotations} we find a equation for the circular geodesic orbits (rotation curves) and study the stability of such orbits in Section \ref{sec:stability} both by the Rayleigh criterion and by a perturbative method. Results and graphics for stable rotation curves and comparison to observed rotation curves of spiral galaxies are showed in Section \ref{sec:curves}. In Section \ref{solar} we introduce quantitative arguments to show that at solar system scales, the model presents the correct behavior for rotation curves (a Newtonian-like profile). In Section \ref{sec:lensing} we derive the gravitational lensing regime of the model to a spherically symmetric cluster of galaxies. Discussions about such results is done in the Concluding Remarks (Section \ref{sec:discussion}). And although the main purpose of the present work is to construct a top-down model and not to shed light on a fundamental microscopic theory we also illustrate another possibility, considering that our model could constrain a Kaluza-Klein dark matter particle to be tested at Large Hadron Collider (LHC) in next years, where we write considerations about the compactification of a $6D$ space-time following the example presented. In what follows, we do $c=G=1$.

\section{Thin disks in $4D$}\label{sec:4ddisks}
Much endeavor has been strongly attached to advance in techniques for finding general relativistic exact solutions in four dimensions \cite{kramer, wald}. One of the earliest accomplishments in this direction was attained by Weyl \cite{weyl}, who found the general static axisymmetric solution of the vacuum Einstein equations:
\be
\label{4dweylmetric}
 \mathrm{d}s^2 = -e^{2\Phi} \mathrm{d}t^2 + e^{-2\Phi} \left( e^{2 \gamma}(\mathrm{d}r^2 + \mathrm{d}z^2) + r^2
 \mathrm{d}\varphi^2 \right),
\ee
where $\Phi(r,z)$ is an arbitrary axisymmetric solution of Laplace's equation in a three-dimensional {\it flat} space with line element
\be\label{flatspace}
 \mathrm{d}\sigma^2 = \mathrm{d}r^2 + r^2 \mathrm{d}\varphi^2 + \mathrm{d}z^2,
\ee
and $\gamma$ satisfies
\be
 \gamma_{,r} = r [\Phi_{,r}^2 - \Phi_{,z}^2],
\ee
\be
\gamma_{,z} = 2 r \Phi_{,r}\Phi_{,z},
\ee where $(\ )_{,a} = \partial /{\partial x^a}$. The solution of these equations is given by a line integral. Since $\Phi$ is harmonic, it can be regarded as a Newtonian potential produced by certain (axisymmetric) sources. Since in this coordinate the spherically symmetric solutions of the Einstein equations correspond to a bar of density 1/2, one needs to be careful in the use of Newtonian images \cite{oliveira}.

As an illustration, the metric (\ref{4dweylmetric}) for a static axially symmetric $4D$ space-time can be written in quasicylindrical coordinates $(r, \varphi, z)$ in the form
\begin{equation}\label{4dmetrica}
\mathrm{d}s^2 = - e^{-\phi} \mathrm{d}t^2 + \chi^2 e^{\phi} \mathrm{d}\varphi^2 +
 f(dr^2 + dz^2),
\end{equation} where $\chi$, $\phi$, and $f$ are functions of $r$ and $z$ only. In the vacuum, the Einstein equations are equivalent to
\begin{equation}\label{eq:lap1}
\chi_{,rr}+\chi_{,zz}=0,
\end{equation}
\begin{equation}\label{eq:lap2}
(\chi\phi_{,r})_{,r}+ (\chi\phi_{,z})_{,z}=0.
\end{equation} Let $\zeta = r + iz$. It is possible to consider $\chi$ as the real part of an analytical function $W(\zeta)=\chi(r,z)+iZ(r,z)$. Noting that $\overline{W}(\zeta)=W(\overline{\zeta})$, one can write $\mathrm{d}W\mathrm{d}\overline{W}=\frac{\partial W}{\partial \zeta}\frac{\partial W}{\partial \overline{\zeta}}\mathrm{d}\zeta \mathrm{d}\overline{\zeta}=|W'(\zeta)|^2 \mathrm{d}\zeta \mathrm{d}\overline{\zeta}$. Or even, $\mathrm{d}W\mathrm{d}\overline{W}=\mathrm{d}\chi^2+\mathrm{d}Z^2=|W'(\zeta)|^2 (\mathrm{d}r^2+\mathrm{d}z^2)$. Thus without lossing generality one can choose $\chi=r$. In such a way we can write $f(r,z)$ as
\begin{eqnarray}\label{eq:lap3}
\ln{f[\phi]}&=&\frac{1}{2}\int r\{[\phi^2_{,r}-\phi^2_{,z}]\mathrm{d}r\nonumber\\
&+& [2\phi_{,r}\phi_{,z}]\mathrm{d}z\}.
\end{eqnarray}

\begin{figure}
\centering
\includegraphics[width=8cm]{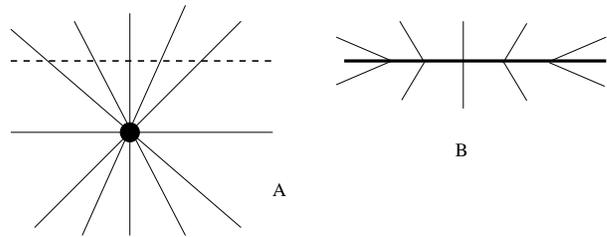}
\caption{Illustration of the ``displace, cut and reflect'' method for the generation of disks.
In A the spacetime with a singularity is displaced and cut by
 a plane (dotted line), in B the part with singularities is disregarded and the
upper part is reflected on the plane.} \label{fig_schem1}
\end{figure}

In order to obtain a solution of (\ref{eq:lap1})--(\ref{eq:lap3}) representing a thin disk located at $z = 0$, we assume that the metric functions $\chi$, $\phi$, and $f$ are continuous across the disk, but have discontinuous first derivatives in the direction normal to the disk. Thin disk solutions in Weyl coordinates are functions of the class $C^0$. The reflectional symmetry of (\ref{eq:lap1}) - (\ref{eq:lap3}) with respect to the plane $z = 0$ allows us to assume that $\chi$, $\phi$, and $f$ are even functions of $z$. Hence, $\chi_{,z}$, $\phi_{,z}$ and $f_{,z}$ are odd functions of $z$. We shall require that they not vanish on the surfaces $z = 0^{\pm}$. Such impositions can induce e.g. a discontinuity  in the space-time by reflecting the solution through the plane. This represents the construction of the disk using the well known ``displace, cut and reflect'' method that was used by Kuzmin \cite{kuzmin} in Newtonian gravity and later in GR by many authors \cite{gonzalez}-\cite{vogtgal}. The  material content of the disk will be described by functions that are distributions with support on the disk. The method can be divided in the following steps that are illustrated in Fig.\ \ref{fig_schem1}. First, in a space wherein we have a compact source of gravitational field, we choose a surface (in our case, the plane $z=0$) that divides the  space in two pieces:  one with no singularities or sources and the other with the sources. Then we disregard the part of the space with singularities and use the surface to make an inversion of the non-singular part of the space. This results in a space with a singularity that is a delta function with support on $z=0$.

This procedure is mathematically equivalent to make the transformation $z \rightarrow |z|+a$, with $a$ constant. In the Einstein tensor we
have first and second derivatives of $z$.  Since $\partial_z |z|=2 \vartheta(z)-1$ and $\partial_{zz} |z|=2\delta(z)$, where $\vartheta(z)$ and $\delta(z)$ are, respectively, the Heaviside  and Dirac distributions, the Einstein field equations will separate in two different pieces \cite{taub}: one valid for $z\not =0$ (the usual Einstein equations), and other involving distributions with an associated energy-momentum tensor. Due to the discontinuous behavior of the derivatives of the metric tensor across the disk, the Riemann curvature tensor contains Dirac delta functions. The energy-momentum tensor can be obtained by the distributional approach due to Papapetrou and Hamouni \cite{papapetrou}, Lichnerowicz \cite{lichnerowicz}, and Taub \cite{taub}. It can be written as ${T^a}_b = [{T^a}_b] \ \delta(z)$, where $\delta$ is the Dirac function with support on the disk and $[{T^a}_b]$ is the distributional energy-momentum tensor, which yield the volume energy density and the principal stresses. A different approach is given by Israel \cite{israel} where one makes use of the extrinsic curvature of the surface to represent the matter. Here, $a,b=0,...,3$.
\subsection{Density and pressures on the disk}
\begin{figure*}
\begin{center}
$\begin{array}{c@{\hspace{0.01in}}c} \multicolumn{1}{l}{\mbox{\bf
(a)}} &
    \multicolumn{1}{l}{\mbox{\bf (b)}} \\ [-0.33cm]
\epsfxsize=3.45in \epsffile{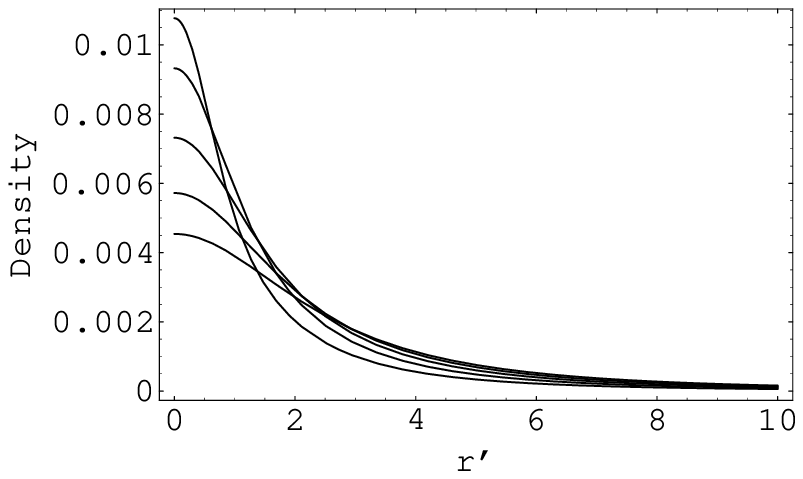} &
    \epsfxsize=3.45in
    \epsffile{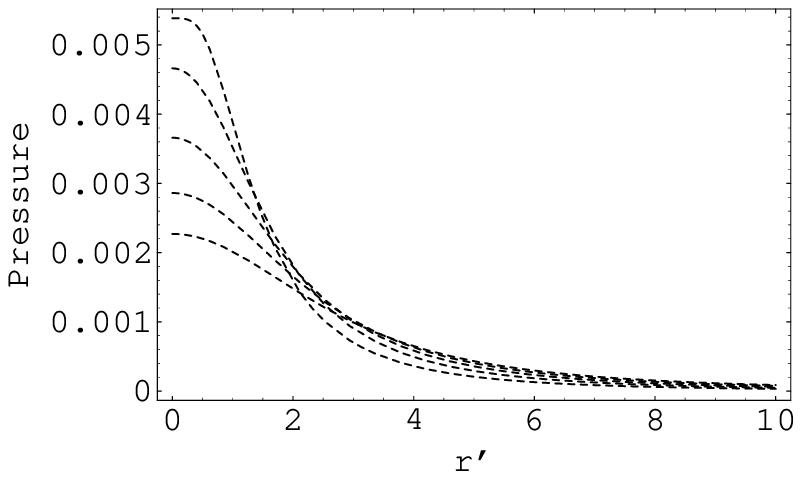} \\ [0.05cm]
\end{array}$
\end{center}
\caption{{\bf (a)} $4D$ disk surface density per unit mass profiles for disk cut
parameters $a=1, 1.5, 2, 2.5, 3$ (from top to bottom). We take $r'=r/m$. {\bf (b)} The same as
{\bf (a)} but for disk pressures. The solution used here is the Schwarzschild one (Chazy-Curzon also offers similar profiles).} \label{fig:4dpressures}
\end{figure*}

The disk at $z=0$ divides the space-time into two halves. The normal to the disk can be described by the co-vector $n_a=\partial z/\partial x^a=(0,0,0,1)$. Above the disk near $z=0$, we can expand the metric as
\be
g_{ab}=g^0_{ab} + z\frac{\partial g_{ab}^+}{\partial z}\vert_{z=0}+z^2\frac{\partial^2 g_{ab}^+}{\partial z^2}\vert_{z=0}+...,
\ee and below $z=0$,
\be
g_{ab}=g^0_{ab} + z\frac{\partial g_{ab}^-}{\partial z}\vert_{z=0}+z^2\frac{\partial^2 g_{ab}^-}{\partial z^2}\vert_{z=0}+....
\ee The quantity $g^0_{ab}$ means the value of $g_{ab}$ at $z=0$. The discontinuities in the first derivatives of the metric tensor can
be cast as
\be \label{discont}
b_{ab} \ = \ g_{ab,z}|_{_{z=0^+}} \ - \ g_{ab,z}|_{_{z = 0^-}}.
\ee The symmetry of the problem gives $\phi_{,z}^+\vert_{_{z=0}}=-\phi_{,z}^-\vert_{_{z=0}}$,  and $f_{,z}^+\vert_{_{z=0}}=-f_{,z}^-\vert_{_{z=0}}$. Denoting $\phi_{,z}\vert_{_{z=0}}=\phi_{,z}^+\vert_{_{z=0}}$, and the same for the other functions, we calculate the following discontinuities:
\begin{equation}
b_{tt}=2\mathrm{e}^{-\phi}\phi_{,z}\vert_{_{z=0}},
\end{equation}
\begin{equation}
b_{rr}=2f_{,z}\vert_{_{z=0}},
\end{equation}
\begin{equation}
b_{zz}=b_{rr},
\end{equation}
\begin{equation}
b_{\varphi \varphi}=2r^2\mathrm{e}^{\phi}\phi_{,z}\vert_{_{z=0}}.
\end{equation}
\noindent It is possible to find the other discontinuities terms by
contracting indices in $b^{ab}$:
\begin{equation}
b^{tt}=-2\mathrm{e}^\phi\phi_{,z}\vert_{_{z=0}}\quad ;\quad
{b^t}_{t}=2\phi_{,z},
\end{equation}
\begin{equation}
b^{rr}=-\frac{2}{f^2}f_{,z}\vert_{_{z=0}}\quad ; \quad
{b^r}_{r}=-\frac{2}{f}f_{,z},
\end{equation}
\begin{equation}
b^{zz}=-\frac{2}{f^2}f_{,z}\vert_{_{z=0}}\quad ; \quad
{b^z}_{z}=-\frac{2}{f}f_{,z},
\end{equation}
\begin{equation}
b^{\varphi
\varphi}=-\frac{2}{r^2}\mathrm{e}^{-\phi}\phi_{,z}\vert_{_{z=0}}\quad ; \quad
{b^\varphi}_{\varphi}=-2\phi_{,z}.
\end{equation} From (\ref{discont}), one can work out (the discontinuities of) the Christoffel symbols through the disk given by
\be
\left[ {\Gamma^a}_{bc} \right] = \frac{1}{2}({b^a}_c{\delta^z}_{b} + {b^a}_b{\delta^z}_{c} - g^{az}b_{bc})
\ee where $\left[ {\Gamma^a}_{bc} \right] \equiv {\Gamma^{+a}}_{bc} - {\Gamma^{-a}}_{bc}$ at $z=0$. From the Riemann tensor defined by
\begin{eqnarray}
R_{abcd}=&&\frac{1}{2}(g_{ad,bc}-g_{bd,ac}+g_{bc,ad}-g_{ac,bd})\nonumber\\
&&+g_{rs}{\Gamma^{r}}_{ad}{\Gamma^{s}}_{bc} - g_{rs}{\Gamma^{r}}_{ac}{\Gamma^{s}}_{bd},
\end{eqnarray} we can compute the Riemann distributional tensor,
\be
\left[ {R^a}_{bcd} \right]=\frac{1}{2}({b^a}_d{\delta^z}_{b}{\delta^z}_{c} - {b^a}_c{\delta^z}_{b}{\delta^z}_{d}+g^{az}b_{bd}{\delta^z}_ c).\ee Defining the Ricci distributional tensor as $[R_{ab}]=[{R^c}_{acb}]$ and the Ricci distributional scalar $[R]=[{R^a}_a]$, we can identify the distributional energy-momentum tensor on the disk through Einstein equations as
\be
[{R^a}_{b}] - \frac{1}{2}{\delta^a}_b[R]=8\pi [{T^a}_b].
\ee Then the distributional energy-momentum tensor is given by
\begin{eqnarray}\label{energy}
[{T^a}_{b}]&&=\frac{1}{16\pi}\{b^{az}{\delta^z}_{b}-b^{zz}\delta^{a}_{b}+g^{az}{b^{z}}_b-g^{zz}{b^{a}}_b\nonumber\\ &&+ {b^{c}}_c(g^{zz}{\delta^{c}}_{b}-g^{az}{\delta^{z}}_{b})\}.\end{eqnarray}
\noindent
This energy-momentum tensor describes the matter content (fluid) of  a thin disk located on $z=0$. The components of a such tensor are calculated as
\begin{equation}
[{T^t}_{t}]=\frac{1}{16\pi}\{-b^{zz}+g^{zz}({b^r}_{r}+{b^z}_{z})\},
\end{equation}
\begin{equation}
[{T^r}_{r}]=\frac{1}{16\pi}\{-b^{zz}+g^{zz}({b^t}_{t}+{b^\varphi}_{\varphi})\},
\end{equation}
\begin{equation}
[{T^z}_{z}]=0,
\end{equation}
\begin{equation}
[{T^\varphi}_{\varphi}]=\frac{1}{16\pi}\{-b^{zz}+g^{zz}({b^t}_{t}+{b^r}_{r})\}.
\end{equation} Defining the vierbein
\begin{eqnarray}
&&{e_{(t)}}^a=\left(\frac{1}{\sqrt{-g_{tt}}},0,0,0\right), {e_{(r)}}^a=\left(0,\frac{1}{\sqrt{g_{rr}}},0,0\right),\nonumber\\
&&{e_{(\varphi)}}^a=\left(0,0,\frac{1}{\sqrt{g_{\varphi \varphi}}},0\right), {e_{(z)}}^a=\left(0,0,0,\frac{1}{\sqrt{g_{zz}}}\right),\nonumber
\end{eqnarray} one can write down the energy-momentum tensor (\ref{energy}) as
\begin{eqnarray}
[{T^a}_b]=&-&\epsilon {e_{(t)}}^a{e_{(t)}}^b + p_r {e_{(r)}}^a{e_{(r)}}^b\nonumber\\
&+&p_\varphi{e_{(\varphi)}}^a{e_{(\varphi)}}^b+p_z{e_{(z)}}^a{e_{(z)}}^b,
\end{eqnarray} yielding the volume densities, i.e., the energy density and pressures as
\be \epsilon = -[{T^t}_t] = -\frac{f_{,z}}{8\pi
f^2}\vert_{_{z=0}}\ee \be p_\varphi = [{T^\varphi}_{\varphi}] =
-\frac{\phi_{,z}}{8\pi f}\vert_{_{z=0}}\ee \be p_r = [{T^r}_r] = 0,
\ee \be p_z = [{T^z}_z] = 0. \ee As an example, in Fig. \ref{fig:4dpressures} we
show the density $\epsilon$ and pressure $p_{\varphi}$ profiles of
the $4D$ constructed thin disk for a Schwarzschild solution (as explained bellow), see also \cite{bicak}.
\subsection{Solutions}\label{subsec:solutions}
In Eq. (\ref{4dmetrica}), the function $\phi$ is intrinsically related to the Newtonian potential $U$ by $\phi=2U$. A important property of Weyl's metric is the fact that Eq. (\ref{eq:lap2}) to be a Laplace's equation in cylindrical coordinates, and exploiting the characteristics of its linearity it is possible to employ the superposition of solutions. Some of the most commonly used asymptotically plane solutions are, e.g.,  Chazy-Curzon, the  finite rod, and the  Schwarzschild  ones.

In the Chazy-Curzon case, the solution for a particle of mass $m$ in the position $z=z_0$ is given by \cite{chazy,curzon}
\be
\phi = \frac{2m}{R}, \qquad \ln f = \frac{m^2 r^2}{R^4},
\ee where $R=\sqrt{r^2+(z-z_0)^2}$.

The Schwarzschild solution corresponds to taking the source for $\phi$ to be a thin rod on the $z$-axis with $1/2$ linear mass density. 
\subsection{Rotation curves}
In first approximation one can consider that  the particles of  such above fluid move along geodesics. In particular, we can  consider particles moving along circular geodesics  whose tangential velocities give us the rotation curves.

\begin{figure}
\centering
\includegraphics[width=8cm]{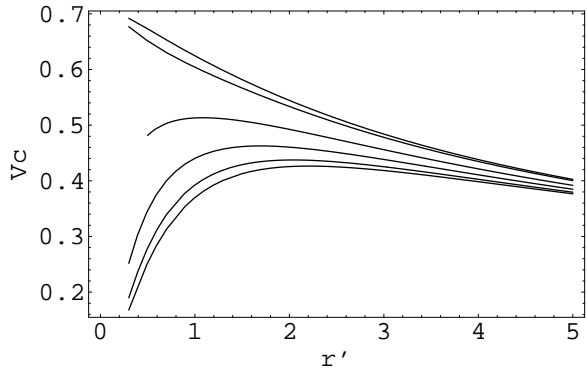}
  \caption{\small $4D$ disk rotation curves  with cut parameters from $a=0.3$ to $a=2$ (from top to bottom). These curves have a Newtonian-like profile. We take $r'=r/m$. The solution used is the Schwarzschild one (Chazy-Curson also offers similar profiles).}
\label{fig:4dcurves}
\end{figure}

From Eq. (\ref{4dmetrica}) we have the first integral of motion,
\begin{equation}\label{4dmetric2}
-\mathrm{e}^{-\phi}\dot{t}^2 + f(\dot{r}^2 + \dot{z}^2) + r^2\mathrm{e}^{\phi}\dot{\varphi}^2 = 1,
\end{equation}
\noindent
where $\dot{x}^a=\mathrm{d}x^a/\mathrm{d}s$. Assuming $\dot{r}=0$ and $\dot{z}=0$ (particles with no radial motion and confined on  $z=0$), Eq.(\ref{4dmetric2})
 reads
\begin{equation}\label{4dlagrange}
-\mathrm{e}^{-\phi}\dot{t}^2 + r^2\mathrm{e}^{\phi}\dot{\varphi}^2 = 1.
\end{equation}
\noindent
The geodesic equations on the disk reduce to
\be
(\mathrm{e}^{-\phi})_{,r}\dot{t}^2 - (r^2\mathrm{e}^{\phi})_{,r}\dot{\varphi}^2=0 \label{4dddr}.
\ee
Eqs. (\ref{4dlagrange}) and (\ref{4dddr}) form a system of equations for  $\dot{\varphi}^2$ and $\dot{t}^2$. From these equations we find the rotation curves $V_C$,
\begin{equation}
V_C=\sqrt{-\frac{g_{\varphi \varphi}}{g_{tt}}}\frac{\mathrm{d}\varphi}{\mathrm{d}t}=\sqrt{-\frac{g_{\varphi \varphi}}{g_{tt}}\frac{\dot{\varphi}^2}{\dot{t}^2}},\label{4dvc1}
\end{equation}
reduce to
\be\label{rocu}
V_C=\sqrt{-\phi_{,r}/ (2/r +\phi_{,r})}.
\ee In Fig. \ref{fig:4dcurves} we show some rotation curves for the constructed $4D$ disk, where we are varying the cut disk parameter $a$. Here we are applying the thin rod solution with linear mass density of 1/2 (Schwarzschild solution).
\section{Axially symmetric solutions for any $D$ and the disk construction}
\subsection{Weyl's geometry for $D$ dimensions}\label{sec:weylD}
The first action to generalize Weyl's geometry to space-times endowed with higher dimensions is to consider a suitable coordinate chart for the  $D$-dimensional line element. Here we follow the method presented by \cite{emparan}, a simple generalization of what is done in four dimensions \cite{wald}. It will be alleged that the metric is Riemannian or Lorentzian. Let $\xi_{(i)}$ represent the Killing vector fields, $1\le i \le D-2$. It is admissible to elect coordinates $(x^i,y^1,y^2)$ such that $\xi_{(i)} = \partial/\partial x^i$, since these commute, with the metric coefficients depending only on $y^1$ and $y^2$.

Now we must show that one can select coordinates $y^1$ and $y^2$ to span two-dimensional surfaces orthogonal to all of the $\xi_{(i)}$. In order to perform this, one has to make evident that the two-dimensional subspaces of the tangent space orthogonal to all of the vectors $\xi_{(i)}$ are integrable. Sufficient conditions for integrability are afforded by the Emparan-Reall theorem \cite{emparan}. If it is further assumed that the Killing vector fields are orthogonal to each other then the metric must take the form 
\be\label{lineelement}
\mathrm{d}s^2 = \sum_{i=1}^{D-2} \epsilon_i e^{2\Phi_i} (\mathrm{d}x^i)^2 + g_{ab} \mathrm{d}y^a \mathrm{d}y^b,
\ee
where $a$ and $b$ take the values $1,2$, the metric coefficients are independent of $x^i$, and $\epsilon_i = \pm 1$ according to whether $\xi_{(i)}$ is spacelike or timelike. Locally it is always possible to find coordinates such that
\be\label{substit}
 g_{ab} \mathrm{d}y^a \mathrm{d}y^b = e^{2C} \mathrm{d}Z \mathrm{d}\bar{Z},
\ee
where $Z$ and $\bar{Z}$ are complex conjugate coordinates if the transverse space is spacelike. The function $C$ is independent of $x^i$. Now, assuming that $\mu, \nu=i,Z,\bar{Z}$, then the vacuum Einstein equations read $R_{\mu\nu} = 0$. The $ij$ component gives
\be
\label{Einsteinij}
 \partial_Z \left[ \exp\left( \sum_j \Phi_j \right) \partial_{\bar{Z}}
 \Phi_i \right] + \partial_{\bar{Z}} \left[ \exp \left( \sum_j \Phi_j
 \right) \partial_Z \Phi_i \right] = 0.
\ee
Summing this equation over $i$ yields
\be
 \partial_Z \partial_{\bar{Z}} \exp \left( \sum_j \Phi_j \right) = 0.
\ee
This last has as general solution
\be
\label{wdef}
 \sum_j \Phi_j  = \log \left( w(Z) + \tilde{w}(\bar{Z}) \right),
\ee
where $\tilde{w} = \bar{w}$ if $Z$ and $\bar{Z}$ are complex conjugate. Substituting equation (\ref{wdef}) into equation (\ref{Einsteinij}) yields
\be
 \label{Aieq}
 2 ( w + \tilde{w} ) \partial_Z \partial_{\bar{Z}} \Phi_i + \partial_Z w \partial_{\bar{Z}} \Phi_i + \partial_{\bar{Z}} \tilde{w} \partial_Z \Phi_i = 0.
\ee
If $w$ is non-constant then $R_{ZZ}=0$ can be rearranged to give
\be
 \partial_Z C = \frac{ \sum_i \partial^2_Z \Phi_i}{\sum_i \partial_Z \Phi_i} + \frac{1}{2} \sum_i \partial_Z \Phi_i - \frac{ \sum_{i<j} \partial_Z \Phi_i \partial_Z \Phi_j }{ 2 \sum_i \partial_Z \Phi_i}.
\ee
An akin equation is obtained from $R_{\bar{Z} \bar{Z}} = 0$:
\be
 \partial_{\bar{Z}} C = \frac{ \sum_i \partial^2_{\bar{Z}} \Phi_i}{\sum_i
 \partial_{\bar{Z}}  \Phi_i} + \frac{1}{2} \sum_i \partial_{\bar{Z}} \Phi_i - \frac{ \sum_{i<j} \partial_{\bar{Z}}  \Phi_i \partial_{\bar{Z}} \Phi_j }{ 2 \sum_i \partial_{\bar{Z}} \Phi_i}.
\ee
Integrating the first two terms of these equations and applying equation (\ref{wdef}) we get
\be
\label{eqn:Csol}
 C = \frac{1}{2} \log \left(\partial_Z w \partial_{\bar{Z}} \tilde{w}\right)  + \Xi,
\ee
where
\be
\label{dZxi}
 \partial_Z \Xi = - \frac{w + \tilde{w}}{\partial_Z w} \sum_{i<j} \partial_Z \Phi_i \partial_Z \Phi_j,
\ee
\be
\label{dZbarxi}
 \partial_{\bar{Z}} \Xi = - \frac{w + \tilde{w}}{\partial_{\bar{Z}} \tilde{w}} \sum_{i<j} \partial_{\bar{Z}} \Phi_i \partial_{\bar{Z}} \Phi_j.
\ee
The integrability condition for $\Xi$ is
\be
 \partial_Z \partial_{\bar{Z}} \Xi = \partial_{\bar{Z}} \partial_Z \Xi.
\ee
It is straightforward to check that this equation is indeed satisfied if equations (\ref{wdef}) and (\ref{Aieq}) hold. These equations also confirm that the remaining Einstein equation $R_{Z \bar{Z}} = 0$ is satisfied.

Since $w$ and $\tilde{w}$ have been assumed non-constant, one can perform a coordinate transformation from $Z$ and $\bar{Z}$ to $w(Z)$ and $\tilde{w}(\bar{Z})$ in the similar way as in the $4D$ case mentioned at the beginning of Sec. \ref{sec:4ddisks}. In $4D$, they are referred as ``Weyl's canonical coordinates'' \cite{kramer}. This yields
\be
\label{weylmetrica}
 \mathrm{d}s^2 = \sum_i \epsilon_i e^{2\Phi_i} (\mathrm{d}x^i)^2 + e^{2\;\Xi} \mathrm{d}w \mathrm{d}\tilde{w}.
\ee
This coordinate transformation is conformal. Eqs. (\ref{Aieq}), (\ref{dZxi}) and (\ref{dZbarxi}) are conformally invariant so the transformation just replaces $\partial_Z$ by $\partial \equiv \partial_w$ and $\partial_{\bar{Z}}$ by $\bar{\partial} \equiv \partial_{\tilde{w}}$. Then the solution is determined by the following equations
\be
 \label{wdef2}
 \sum_i \Phi_i = \log ( w + \tilde{w}),
\ee
\be
 \label{Aieq2}
 2 (w + \tilde{w}) \partial \bar{\partial} \Phi_i + \partial \Phi_i + \bar{\partial} \Phi_i = 0,
\ee
\be
 \label{dwxi}
 \partial \Xi = - (w+ \tilde{w}) \sum_{i<j} \partial \Phi_i \partial \Phi_j,
\ee
\be
 \label{dwbarxi}
 \bar{\partial} \Xi = - (w + \tilde{w}) \sum_{i<j} \bar{\partial} \Phi_i \bar{\partial} \Phi_j.
\ee
If $Z$ and $\bar{Z}$ are complex conjugate coordinates then, as mentioned above, one must take $\tilde{w} = \bar{w}$. Introduce real coordinates $(r,z)$ by $w = r + iz$, so the canonical form of the metric is
\be
 \mathrm{d}s^2= \sum_i \epsilon_i e^{2\Phi_i} (\mathrm{d}x^i)^2 + e^{2\;\Xi}(\mathrm{d}r^2+\mathrm{d}z^2).
\ee
Eq. (\ref{Aieq2}) then takes the form
\be
 \frac{\partial^2 \Phi_i}{\partial r^2} + \frac{1}{r}\frac{\partial \Phi_i}{\partial r} + \frac{\partial^2 \Phi_i}{\partial z^2} = 0,
\ee
which is again just Laplace's equation in three-dimensional flat space with metric (\ref{flatspace}). The function $\Phi_i$ is independent of the coordinate $\theta$, i.e., it is axisymmetric. The solution is therefore specified by $D-3$ independent axisymmetric solutions of Laplace's equation in three-dimensional flat space. 
\subsection{The simplest example: a $6D$ disk}\label{sec:6D}
The fact that $\Phi_i$, for $D$ dimensions, has as solution axisymmetric solutions of Laplace's equation leads to the possibility to construct a disk with higher dimensional components. To obtain a solution for such a metric which represents a thin disk located on $y^a=z=0$, we assume the functions of the metric $\Phi_i$ and $g_{ab}$ are continuous along the disk, in particular on the surface $z=0$, but with discontinuous first derivatives on that surface. Following the same approach developed in Sec. \ref{sec:4ddisks}, we can  introduce these discontinuities by doing the replacement $z\rightarrow |z|+ a$, where $a$ is a constant. And as in Sec. \ref{sec:4ddisks}, such discontinuities indicate that the Riemann curvature tensor contains Dirac delta functions that makes possible to calculate the energy-momentum tensor by a distributional approach (see also Sec. \ref{sec:4ddisks} for a full explanation about the method). Calculating the extradimensional components of the energy-momentum tensor, we find that $[T^{x^i}_{x^i}]=-[T^{x^{i+1}}_{x^{i+1}}]=\frac{1}{16\pi}g^{zz}b^{x^i}_{x^i}$. If $D$ is even, then the sum of extradimensional pressures cancel and the total pressure depends only of $4D$ components. Thus we assume a disk living in universe endowed with an even number of extra dimensions. Other important arguments can be found in literature in favor of an even $D$ (e.g. as has been emphasized by different authors, the Huygens principle does not hold for odd $D$ \cite{cardoso}).

Whatsoever, as the simplest example, we can work out in what follows the case for a $6D$ disk. The metric for an axially symmetric 6D space-time can be written in quasi-cylindrical coordinates as
\begin{equation}\label{metrica}
\mathrm{d}s^2 = -\mathrm{e}^{-\phi}\mathrm{d}t^2  + \chi^2\mathrm{e}^{\phi}\mathrm{d}\varphi^2+\psi\mathrm{e}^\nu\mathrm{d}x^2 + \mathrm{e}^{-\nu}\mathrm{d}y^2 + f(\mathrm{d}r^2 + \mathrm{d}z^2),
\end{equation}
\noindent
where $\phi=\phi(r,z)$, $f=f(r,z)$, $\chi=\chi(r,z)$, $\psi = \psi(r,z)$
 and $x$ and $y$ are the extra dimensional coordinates.  The vacuum Einstein equations $R_{AB}=0$, ($A,B=0,1,...,5$) reduce to
\begin{equation}\label{h}
(\chi\sqrt{\psi})_{,rr}+(\chi\sqrt{\psi})_{,zz}=0,
\end{equation}
\begin{equation}\label{lap1}
\nu_{,rr}+ \frac{\nu_{,r}(\chi\sqrt{\psi})_{,r}}{\chi\sqrt{\psi}}+\frac{\nu_{,z}(\chi\sqrt{\psi})_{,z}}{\chi\sqrt{\psi}}+\nu_{,zz}=0,
\end{equation}
\begin{equation}\label{lap2}
\phi_{,rr}+ \frac{\phi_{,r}(\chi\sqrt{\psi})_{,r}}{\chi\sqrt{\psi}}+\frac{\phi_{,z}(\chi\sqrt{\psi})_{,z}}{\chi\sqrt{\psi}}+\phi_{,zz}=0.
\end{equation}
\noindent
where $(\;)_{,a} = \partial/\partial x^a$. About the coordinate freedom, one can consider $\varsigma=\chi\sqrt{\psi}$ as the real part of an analytical function $W(\zeta)=\varsigma(r,z)+iZ(r,z)$, where $\zeta = r + iz$. Noting that $\overline{W}(\zeta)=W(\overline{\zeta})$, one can write $\mathrm{d}W\mathrm{d}\overline{W}=\frac{\partial W}{\partial \zeta}\frac{\partial W}{\partial \overline{\zeta}}\mathrm{d}\zeta \mathrm{d}\overline{\zeta}=|W'(\zeta)|^2 \mathrm{d}\zeta \mathrm{d}\overline{\zeta}$. Or even, $\mathrm{d}W\mathrm{d}\overline{W}=\mathrm{d}\varsigma^2+\mathrm{d}Z^2=|W'(\zeta)|^2 (\mathrm{d}r^2+\mathrm{d}z^2)$. Thus without lossing generality one can choose $\chi\sqrt{\psi}=r$. We do  $\psi=1$ and $\chi=r$, that  yields the  Einstein equation for $f(r,z),$
\begin{eqnarray}\label{f}
\ln{f[\phi,\nu]}&=&\frac{1}{2}\int
r\{[\phi^2_{,r}-\phi^2_{,z}+2\phi_{,r}/r+\nu^2_{,r}-\nu^2_{,z}]\mathrm{d}r\nonumber\\ &+& [2\phi_{,r}\phi_{,z}+2\phi_{,z}/r+2\nu_{,r}\nu_{,z}]\mathrm{d}z\}.
\end{eqnarray}
\noindent
Eqs.(\ref{h})--(\ref{f}) form the complete set of vacuum Einstein equations for the metric (\ref{metrica}).

To obtain a solution of (\ref{h})--(\ref{f}) which represents a thin disk located on $z=0$, we assume the functions of the metric $f$ and $\phi$ are continuous along the disk, in particular on the surface $z=0$, but with discontinuous first derivatives on that surface. We introduce these discontinuities by doing the
replacement $z\rightarrow |z|+ a$, where $a$ is a constant. The distributional energy-momentum tensor reads
\begin{equation}
[T^t\;_{t}]=\frac{1}{16\pi}\{-b^{zz}+g^{zz}(b^r\;_{r}+b^z\;_{z}+b^\varphi\;_{\varphi}+b^x\;_{x}+b^y\;_{y})\},
\end{equation}
\begin{equation}
[T^r\;_{r}]=\frac{1}{16\pi}\{-b^{zz}+g^{zz}(b^t\;_{t}+b^z\;_{z}+b^\varphi\;_{\varphi}+b^x\;_{x}+b^y\;_{y})\},
\end{equation}
\begin{equation}
[T^z\;_{z}]=0,
\end{equation}
\begin{equation}
[T^\varphi\;_{\varphi}]=\frac{1}{16\pi}\{-b^{zz}+g^{zz}(b^t\;_{t}+b^r\;_{r}+b^z\;_{z}+b^x\;_{x}+b^y\;_{y})\},
\end{equation}
\begin{equation}
[T^x\;_{x}]=\frac{1}{16\pi}\{-b^{zz}+g^{zz}(b^t\;_{t}+b^r\;_{r}+b^z\;_{z}+b^\varphi\;_{\varphi}+b^y\;_{y})\},
\end{equation}
\begin{equation}
[T^y\;_{y}]=\frac{1}{16\pi}\{-b^{zz}+g^{zz}(b^t\;_{t}+b^r\;_{r}+b^z\;_{z}+b^\varphi\;_{\varphi}+b^x\;_{x})\}.
\end{equation} And the discontinuities of first derivatives of the metric $b_{AB}=(g_{AB,z}^+ - g_{AB,z}^-)\vert_{_{z=0}}$ yields
\begin{equation}
b_{tt}=2\mathrm{e}^{-\phi}\phi_{,z}\vert_{z=0},
\end{equation}
\begin{equation}
b_{rr}=2f_{,z}\vert_{z=0},
\end{equation}
\begin{equation}
b_{zz}=b_{rr},
\end{equation}
\begin{equation}
b_{\phi \phi}=2r^2\mathrm{e}^{\phi}\phi_{,z}\vert_{z=0},
\end{equation}
\begin{equation}
b_{xx}=2\mathrm{e}^{\nu}\nu_{,z}\vert_{z=0},
\end{equation}
\begin{equation}
b_{yy}=-2\mathrm{e}^{-\nu}\nu_{,z}\vert_{z=0}.
\end{equation} 
Including the extra dimensions  the vierbein is replaced by the 
sechsbein,

\begin{widetext}
\ba
&&{e_{(t)}}^A=\left(\frac{1}{\sqrt{-g_{tt}}},0,0,0,0,0\right), {e_{(r)}}^A=\left(0,\frac{1}{\sqrt{g_{rr}}},0,0,0,0\right),{e_{(\varphi)}}^A=\left(0,0,\frac{1}{\sqrt{g_{\varphi \varphi}}},0,0,0\right),\nonumber\\ &&{e_{(z)}}^A=\left(0,0,0,\frac{1}{\sqrt{g_{zz}}},0,0\right),{e_{(x)}}^A=\left(0,0,0,0,\frac{1}{\sqrt{g_{xx}}},0\right), {e_{(y)}}^A=\left(0,0,0,0,0,\frac{1}{\sqrt{g_{y y}}}\right),\nonumber
\ea
\end{widetext} 
\noindent one can write down the energy-momentum tensor (\ref{energy}) as
\begin{eqnarray}
[{T^A}_B]=&-&\epsilon {e_{(t)}}^A{e_{(t)}}^B + p_r {e_{(r)}}^A{e_{(r)}}^B\nonumber\\
&+&p_\varphi{e_{(\varphi)}}^A{e_{(\varphi)}}^B+p_z{e_{(z)}}^A{e_{(z)}}^B\nonumber\\
&+&p_x{e_{(x)}}^A{e_{(x)}}^B+p_y{e_{(y)}}^A{e_{(y)}}^B,
\end{eqnarray} yielding the energy density and pressures as
\be\label{p1} \epsilon = -[{T^t}_t] = -\frac{f_{,z}}{8\pi
f^2}\vert_{_{z=0}}\ee \be\label{p2} p_\varphi = [{T^\varphi}_{\varphi}] =
-\frac{\phi_{,z}}{8\pi f}\vert_{_{z=0}}\ee \be\label{p3} p_r = [{T^r}_r] = 0,
\ee \be\label{p4} p_z = [{T^z}_z] = 0. \ee  \be\label{p5} p_x+p_y = 0. \ee 

A second step to construct the disk  is to  choose
 two  solutions of Laplace equations (\ref{lap1}) and (\ref{lap2}) for the functions $\phi$ and  $\nu$. Usually, for thin disks in $4D$ it is common to solve the problem with Schwarzschild or Chazy-Curzon solutions \cite{bicak} (see Sec. \ref{subsec:solutions}). The $D$-dimensional Schwarzschild solution has isometry group ${\bf R}\times O(D-1)$. To write it in Weyl form, $D-2$ orthogonal commuting Killing vector fields are required. For the Schwarzschild solution, this occurs only for $D=4,5$. Hence only the four and five-dimensional Schwarzschild solutions can be written in Weyl form. For $D>5$, the $D$-dimensional Schwarzschild solution is not a generalized Weyl solution. However, the geometry obtained by taking products of the $D=4$ or $D=5$ Schwarzschild solution with asymptotically flat space are easily seen to be Weyl solutions. 

We take the Newtonian potential associated to  a rod  of constant density \cite{schwarzschild},
\begin{equation}\label{barra}
\phi(r,z)=\ln{\frac{R_1 + R_2 - 2}{R_1+R_2+2}},
\end{equation}
\noindent
where $R_1=\sqrt{\tilde{r}^2+(z-1)^2}$ e $R_2=\sqrt{\tilde{r}^2+(z+1)^2}$, and
$\tilde{r}$ is the radial coordinate normalized by mass, $\tilde{r}=r/m$. The length of the rod is $L=2m$. We also use a Chazy-Curzon solution for the $\nu$ function ($\nu(r,z)= [r^2 + (|z|+a)^2]^{-1/2}$), which is asymptotically flat, resulting that our disk is in fact a Weyl solution. Thus, along the rest of the present work, the $4D$ part is solved using the Schwarzschild solution, or the Chazy-Curzon, and the extradimensional part by using the Chazy-Curzon solution. Both pictures produce very similar results.
\section{Rotation curves}\label{sec:rotations}
From Eq. (\ref{metrica}) we have the first integral of motion,
\begin{equation}\label{metric2}
-\mathrm{e}^{-\phi}\dot{t}^2 + f(\dot{r}^2 + \dot{z}^2) + r^2\mathrm{e}^{\phi}\dot{\varphi}^2+\mathrm{e}^\nu\dot{x}^2 + \mathrm{e}^{-\nu}\dot{y}^2 = 1,
\end{equation}
\noindent
where $\dot{x}^A=\mathrm{d}x^A/\mathrm{d}s$. Assuming $\dot{r}=0$ and $\dot{z}=0$ (particles with no radial motion and confined on  $z=0$), Eq.(\ref{metric2})
 reads
\begin{equation}\label{lagrange}
-\mathrm{e}^{-\phi}\dot{t}^2 + r^2\mathrm{e}^{\phi}\dot{\varphi}^2+\mathrm{e}^\nu\dot{x}^2 + \mathrm{e}^{-\nu}\dot{y}^2 = 1.
\end{equation}
\noindent
By Eqs. (\ref{p1})--(\ref{p5}), one evidently can see that extra dimensions pressures do not contribute to total disk pressures. Immediately one is tempted to argue that this is the same to say that extra dimensions do not contribute to the density profiles. However, Eq. (\ref{f}) shows that $f(r,z)$ contains extra dimensional field components and therefore {\it a priori} the rates $\dot{x}$ and $\dot{y}$ must be considered as non-null incognites. On this fashion, the geodesic equations on the disk reduce to
\begin{eqnarray}
&&\mathrm{e}^\nu \dot{x} = C_x, \;\;\; \mathrm{e}^{-\nu} \dot{y} = C_y, \label{cxcy}\\
&&(\mathrm{e}^{-\phi})_{,r}\dot{t}^2 - (r^2\mathrm{e}^{\phi})_{,r}\dot{\varphi}^2\nonumber\\&=&C_x^2(\mathrm{e}^{-\nu})_{,r}+C_y^2(\mathrm{e}^{\nu})_{,r} \label{ddr},
\end{eqnarray}
where $C_x$ and $C_y$ are integration constants.

Eqs. (\ref{lagrange}) and (\ref{ddr}) form a system of equations for  $\dot{\varphi}^2$ and $\dot{t}^2$. From these equations we find the rotation curves $V_C$,
\begin{equation}
V_C=\sqrt{-\frac{g_{\varphi \varphi}}{g_{tt}}}\frac{\mathrm{d}\varphi}{\mathrm{d}t}=\sqrt{-\frac{g_{\varphi \varphi}}{g_{tt}}\frac{\dot{\varphi}^2}{\dot{t}^2}},\label{vc1}
\end{equation}
reduce to
\begin{equation}
V_C=\sqrt{\frac{F(r)\nu_{,r} + G(r)\phi_{,r}}{ F(r)\nu_{,r} - G(r)(2/r+\phi_{,r})}}, \label{vc2}
\end{equation}
\noindent
where $F(r)=-C^{2}_x \mathrm{e}^{-\nu} +C^{2}_y \mathrm{e}^{\nu}$ and $G(r)=1- C^{2}_x \mathrm{e}^{-\nu} -C^{2}_y e^{\nu}$. Note that when $C_x=C_y =0$ (no extra dimensions), we have $V_C=\sqrt{-\phi_{,r}/ (2/r +\phi_{,r})}$ that is the known formula for circular orbits in 4D Weyl geometry, Eq. (\ref{rocu}). Another important task about the derived rotation curves is that Eqs. (\ref{cxcy}) and (\ref{ddr}) provide that the disk actually is living in a universe endowed with Universal Extra Dimensions (a model where the fields can access the extra dimensions; for details about the model see \cite{antoniadis,acd,kkdm}). The boundary conditions in manner to determine values for $C_x$ and $C_y$ come from calculations of the stability of the disk, a subject studied in what follows.
\section{The stability}\label{sec:stability}
We need to answer the question: what are the values for $C_x$ and $C_y$ (the integration constants) that make the disk stable? A classical method, the Rayleigh criterion, and an advanced one, a pertubative method, are both used to constrain a possible range of stable values for $C_x$ and $C_y$. The correct rotation curves must be constructed following the results from a such stability.
\subsection{Rayleigh criterion}
\begin{figure*}
\begin{center}
$\begin{array}{c@{\hspace{0.01in}}c} \multicolumn{1}{l}{\mbox{\bf
(a)}} &
    \multicolumn{1}{l}{\mbox{\bf (b)}} \\ [-0.33cm]
\epsfxsize=3.45in \epsffile{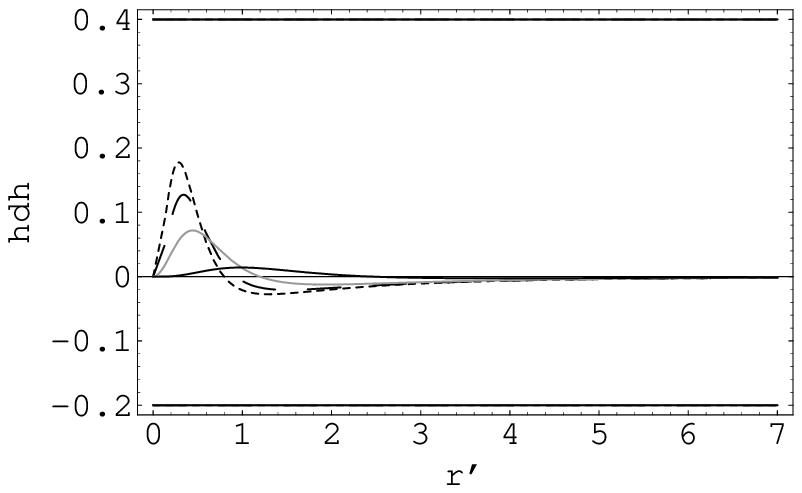} &
    \epsfxsize=3.45in
    \epsffile{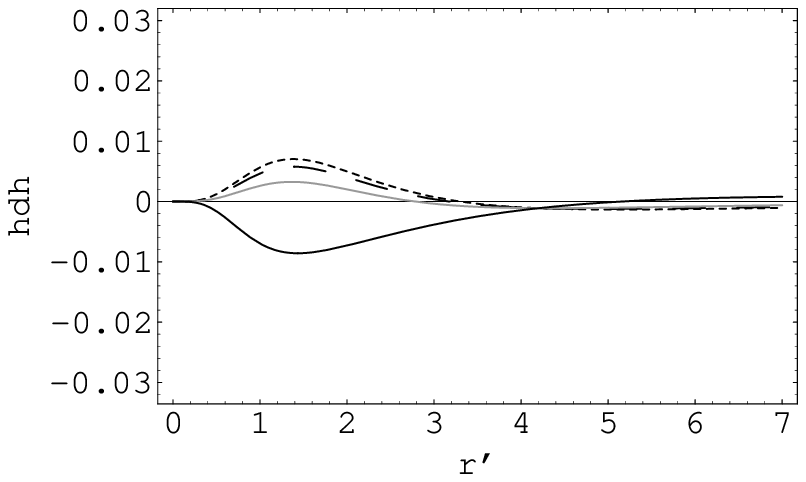} \\ [0.05cm]
\end{array}$
\hspace{0.01in}\mbox{\bf (c)}\\\epsfxsize=3.45in
    \epsffile{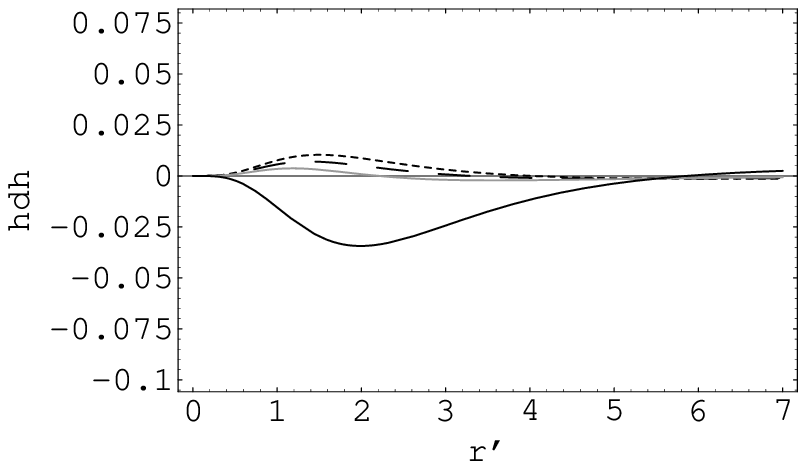}
\end{center}
\caption{Stability of disks by the modified Rayleigh criterion. In {\bf (a)} we show that stable disk cut parameters $a$ are obtained only for $a>1$ (the full line), where $h\frac{\mathrm{d}h}{\mathrm{d}r}>0$. {\bf (b)} In the region of interest, stable disks occur for extradimensional parameters $C_x=0.8$ (dotted line), $C_x=0.85$ (dashed line) and $C_x=0.9$ (gray line). For $C_x>0.95$, $h\frac{\mathrm{d}h}{\mathrm{d}r}<0$, and the disk becomes unstable (full line). {\bf (c)} In the region of interest, stable disks occur for extradimensional parameters $C_y=0.1$ (dotted line), $C_y=0.2$ (dashed line) and $C_y=0.3$ (gray line). For $C_y>0.5$, $h\frac{\mathrm{d}h}{\mathrm{d}r}<0$, and the disk becomes unstable (full line). Here $a$ is normalized by mass and $r'=r/m$.} \label{fig:rayleigh}
\end{figure*}

The  stability of the circular orbits in the disk plane can  be studied using an extension of the Rayleigh stability criterion \cite{vogt20}. This method is extremely good for Newtonian systems, and only for circular orbits. Given a such limitation, here it will be used only as a reference method. In Rayleigh criterion, we have stability when $h\frac{\mathrm{d}h}{\mathrm{d}r}>0,$ where $h$ is the specific angular momentum of a particle in the disk plane ($h=g_{\varphi \varphi}\dot{\varphi}$),
\begin{equation}
h=r^2\mathrm{e}^{\phi}\sqrt{\frac{(1- C^{2}_x \mathrm{e}^{-\nu} -C^{2}_y \mathrm{e}^{\nu})\phi_{,r}+ (C_x^2\mathrm{e}^{-\nu}-C_y^2\mathrm{e}^{\nu})\nu_{,r}}{2r^2\mathrm{e}^{\phi}(1/r+ \phi_{,r})}}.
\end{equation}
\noindent
We find that  for different values of  $C_x$ and $C_y$ less than unity   stability is reached when  $a>1$.  For small values of $a$ ($a<1$, highly relativistic disks) we have a small zone of instability, typically around $r=3m$. In Fig. \ref{fig:rayleigh} we show the stability of the disk by the criterion presented above. The principal constraint derived from this method is to hold suitable values for the cut parameter $a$.
\subsection{Perturbative method}
The  stability of circular orbits in the disk plane can  be studied using a more acurate and sophisticated method. In what follows we compute a perturbative method where we assume that the disk particles are describing equatorial circular geodesics in stationary axisymmetric fields. The perturbation of the geodesic equation $\ddot{x}^A+{\Gamma^A}_{BC}\dot{x}^B\dot{x}^C=0$ is done performing the transformation $x^A\rightarrow x^A+\Delta^A$ -- where $\Delta^A=(\delta t, \delta r, \delta \varphi, \delta z, \delta x,\delta y)$ are infinitesimal elements. Therefore, equations for the perturbations are,
\begin{equation}\label{perturb}
\ddot{\Delta}^A + 2{\Gamma^A}_{BC}\dot{x}^B\dot{\Delta}^C +
{\Gamma^A}_{BC,D}\Delta^D\dot{x}^B\dot{x}^C=0,
\end{equation}
\noindent where ${\Gamma^A}_{BC}$ are the Christoffel symbols and $\dot{x}^A$ are proper time derivatives $\mathrm{d}x^A/\mathrm{d}s$ and can be written for a circular orbital motion as
\begin{equation}\label{veloc}
\dot{x}^A=(u^t,0,0,u^t\Omega,C_x,C_y),
\end{equation}
\noindent where $u^t\Omega=V_C$. Note that (\ref{perturb}) is equivalent to the usual deviation equation \cite{shirokov}. Assuming only horizontal oscillations in the 4D part of the disk ($\delta z=0$), we get
\begin{equation}
\Delta^{A}=(\delta t, \delta r,0,0,\delta \varphi,0,0).
\end{equation}
\noindent Let $x^A$ be an equatorial circular geodesic in a stationary axisymmetric space-time (\ref{metrica}), i.e., the worldline $x^A=(t, r=\mathrm{const}, \varphi = \mathrm{const} + \Omega t, z=0, x=\mathrm{const}, y=\mathrm{const})$. Substituting the four velocity (\ref{veloc}) and demanding that $g_{AB,z}$ (but not $g_{AB,zz}$) vanishes in the equatorial plane, the components of Eq.(\ref{perturb}) for horizontal oscillations read:
\begin{eqnarray}
&&(\ddot{\delta t}) + 2{\Gamma^t}_{tr}u^t(\dot{\delta r})=0,\label{1}\\
&&(\ddot{\delta r}) + 2{\Gamma^r}_{tt} + 2{\Gamma^r}_{\varphi \varphi}u^t\Omega(\dot{\delta \varphi})+[({\Gamma^r}_{tt,r}\\\nonumber&&+{\Gamma^r}_{\varphi \varphi,r}\Omega^2)(u^t)^2 + {\Gamma^r}_{xx,r}C_x^2 +{\Gamma^r}_{yy,r}C_y^2]\delta r=0,\label{2}\\
&&(\ddot{\delta \varphi})+2{\Gamma^\varphi}_{\varphi r}\Omega u^t
(\dot{\delta r})=0.\label{3}\end{eqnarray}

\noindent Suppose that the solutions for $\delta t$, $\delta r$ and
$\delta \varphi$ have a form of harmonic oscillations, $\sim e^{iKs}$, with a common proper angular frequency $K$. The condition for solvability of Eqs. (\ref{1})-(\ref{3}) is then
\begin{equation}
\det{\begin{pmatrix}
-K^2&2iK{\Gamma^t}_{tr}u^t&0\\
2iK{\Gamma^r}_{tt}u^t&-K^2+{\Gamma^r}_{AB,r}u^Au^B&2iK{\Gamma^r}_{\varphi \varphi}u^t\Omega\\
0&2iK{\Gamma^\varphi}_{\varphi r}u^t\Omega&-K^2
\end{pmatrix}}=0,
\end{equation}
\noindent where ${\Gamma^r}_{AB,r}u^Au^B=({\Gamma^r}_{tt,r}+{\Gamma^r}_{\varphi \varphi,r}\Omega^2)(u^t)^2 + {\Gamma^r}_{xx,r}C_x^2 +{\Gamma^r}_{yy,r}C_y^2$. From the non-trivial solution of this equation we derive the oscillation frequency with respect to infinity $\kappa=K/u^t$, referred in literature as the epicyclic frequency \cite{kato}, as
\begin{equation}\label{kappa2}
\kappa^2=\frac{Z(r)}{2+\phi_{,r}-P(r)}[\phi_{,rr}+r\phi_{r}^3+3\phi_{,r}/r+3\phi_r^2+Q(r)],
\end{equation}
\noindent where $Z(r)=-e^{-\phi}/f$, the metric is that given by Eq.(\ref{metrica}) and $P(r)$ and $Q(r)$ are terms related to extradimensional imprints:
\begin{widetext}
\begin{equation}
P(r)=M(r)[2+\phi_{,r}r]+F(r)r\nu_{,r},
\end{equation}
\begin{equation}
Q(r)=P(r)[H(r)/r+0.5\phi_{,r}f_{,r}/f-\phi^2_{,r}/2-\phi_{,rr}/2]-H(r)\phi_{,r}-2H(r)/r-N(r)/r,
\end{equation}
\begin{equation}
H(r)=\frac{e^{\nu-\phi}}{2r}C_x^2[\nu_{,r}f_{,r}/f-\nu^2_{,r}-\nu_{,rr}]+\frac{e^{-\nu-\phi}}{2r}C_y^2[-\nu_{,r}f_{,r}/f-\nu^2_{,r}+\nu_{,rr}]
\end{equation}
\begin{equation}
N(r)=F(r)[-3\nu_{,r}-2r\phi_{,r}\nu_{,r}-0.5r^2\phi_{,r}^2\nu_{,r}+0.5r^2\phi_{,rr}]+M(r)[3\phi_{,r}+2r\phi_{,r}^2+0.5r^2\phi_{,r}^3-0.5r^2\phi_{,rr}\phi_{,r}+r\phi_{,r}f_{,r}/f+0.5r^2\phi^2_{,r}f_{,r}/f],
\end{equation}
\end{widetext}
\noindent and where $M(r)=C_x^2e^{-\nu}+C_y^2e^{\nu}$ and $F(r)=-C_x^2e^{-\nu}+C_y^2e^{\nu}$. When $C_x=C_y=0 \Rightarrow P(r)=Q(r)=0$ (no extra dimensions), Eq.(\ref{kappa2}) becomes the known formula for oscillations in Weyl geometry.
\begin{figure*}
\begin{center}
$\begin{array}{c@{\hspace{0.01in}}c} \multicolumn{1}{l}{\mbox{\bf
(a)}} &
    \multicolumn{1}{l}{\mbox{\bf (b)}} \\ [-0.33cm]
\epsfxsize=3.30in \epsffile{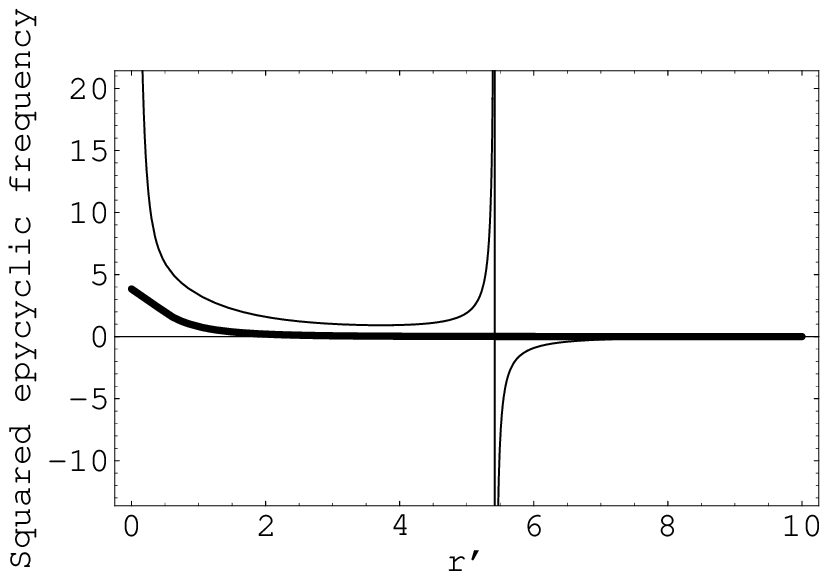} &
\epsfxsize=3.40in
    \epsffile{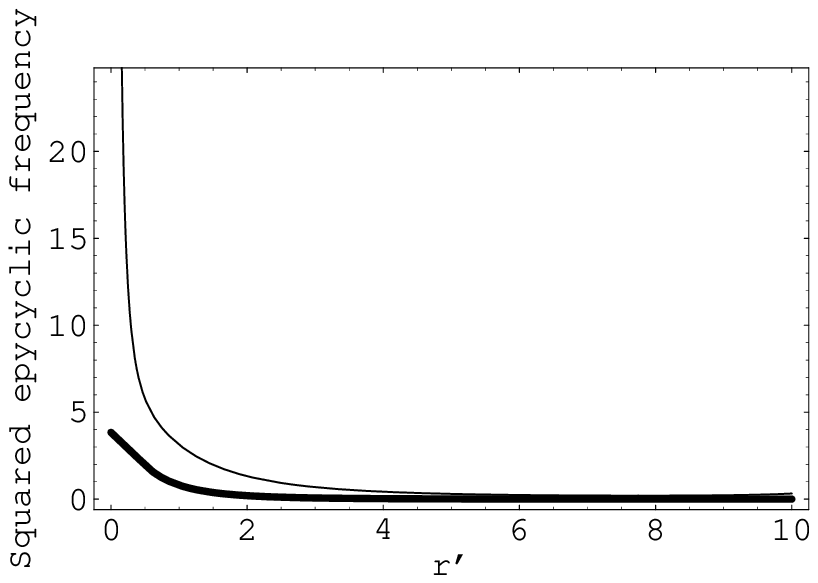} \\ [0.05cm]
\end{array}$
\end{center}
\caption{{\bf (a)} Epicyclic frequency versus the normalized radius $r'=r/m$ for a configuration where $C_x=0.1$ and $C_y=0.85$. The stability is achieved for $0<r'<5.5$, when $\kappa^2>0$; a great number of rotation curves of spiral galaxies remains exactly into this region. The bold curve is the result for a 4D thin disk configuration. {\bf (b)} Epicyclic frequency versus the normalized radius for a configuration where $C_x=0.1$ and $C_y=0.9$. Here the stability is achieved for $0<r'<15$, when $\kappa^2>0$. Here the disk cut parameter $a$ is normalized by mass. We used $r'=r/m$ and $a=1.5$.} \label{fig:epicyclic}
\end{figure*}
\noindent Graphics for the squared epicyclic frequency is showed in Fig. \ref{fig:epicyclic}. The configuration is stable only when $\kappa^2>0$. Our results show that the integration constants $C_x$ and $C_y$ have a very restricted range of stable values. In Table \ref{table1} it is possible to see the intervals for $C_x$ and $C_y$ where the disk is stable (here we fixed $a=1.5$, a stable disk parameter according to the Rayleigh criterion; any values $a>1$ are able to produce stable disks).
\medbreak
\begin{center}
\begin{table}
\caption{Stable values for $C_x$ and $C_y$}
\begin{tabular}{||c|c||}\hline\hline
Values for $C_x$ and $C_y$&Region where the disk is stable\\\hline\hline
$C_x=0$, $C_y=0$ (Newtonian)&{\bf All $r'$}\\\hline
$0<C_x<0.2$, $0<C_y<0.4$&{\bf All $r'$}\\\hline
$0<C_x<0.2$, $C_y=0.5$&$0<r'\lesssim 0.4$\\\hline
$0<C_x<0.2$, $C_y=0.7$&$0<r'\lesssim 1.4$\\\hline
$0<C_x<0.2$, $C_y=0.75$&$0<r'\lesssim 2.6$\\\hline
$0<C_x<0.2$, $C_y=0.8$&$0<r'\lesssim 4$\\\hline
$0<C_x<0.2$, $C_y=0.85$&$0<r'\lesssim 5.5$\\\hline
$0<C_x<0.2$, $C_y=0.9$&$0<r'\lesssim 15$\\\hline
$0<C_x<0.2$, $C_y>0.95$&{\bf Unstable disk}\\\hline
\end{tabular} \label{table1} \end{table}\end{center}\medbreak

A such stability study is fundamental to discuss what are the exact values to be used for the extradimensional
parameters $C_x$ and $C_y$.  In general, extra dimensions contribute to destabilize the disk, but it is possible to derive a semi-phenomenological model where we have a very short range of stable values that can be used to fit rotation curves of galaxies or gravitational lensing of dark halos in a UED background (see Table \ref{table1}). Those stable values can astrophysically constrain two-UED models. A more detailed discussion on the present perturbative method is presented in \cite{letter}.
\subsection{Stable rotation curves}\label{sec:curves}
\begin{figure*}
\begin{center}
$\begin{array}{c@{\hspace{0.01in}}c} \multicolumn{1}{l}{\mbox{\bf
(a)}} &
    \multicolumn{1}{l}{\mbox{\bf (b)}} \\ [-0.33cm]
\epsfxsize=3.45in \epsffile{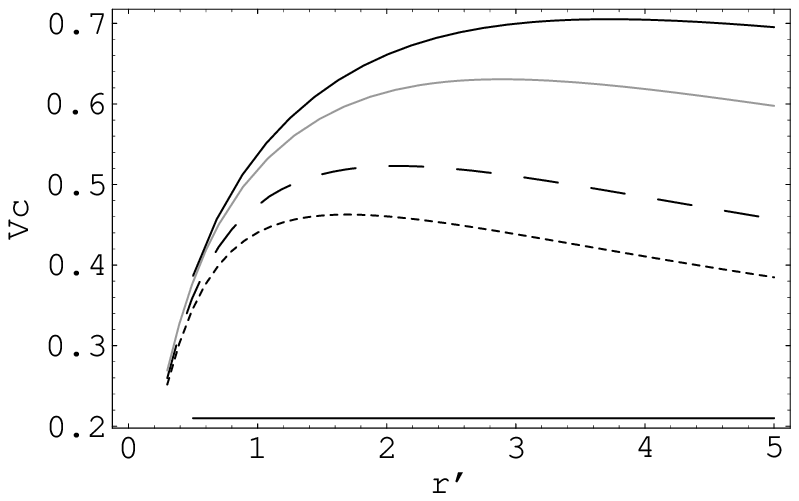} &
    \epsfxsize=3.45in
    \epsffile{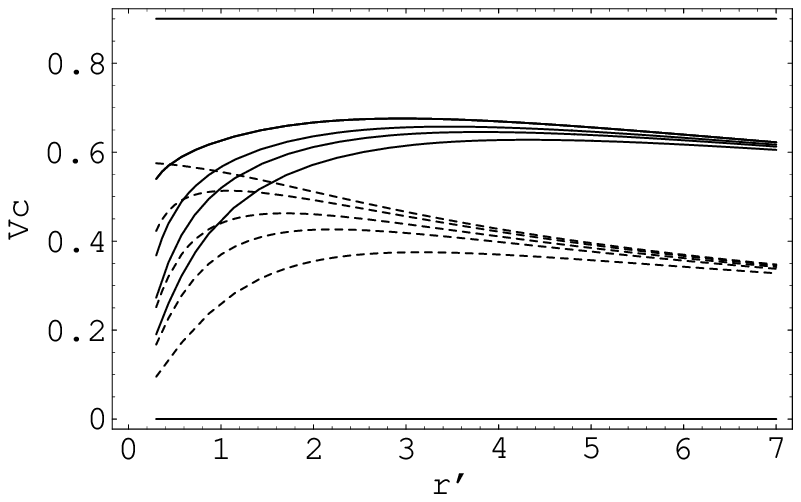} \\ [0.05cm]
\end{array}$
\hspace{0.01in}\mbox{\bf (c)}\\\epsfxsize=3.45in
    \epsffile{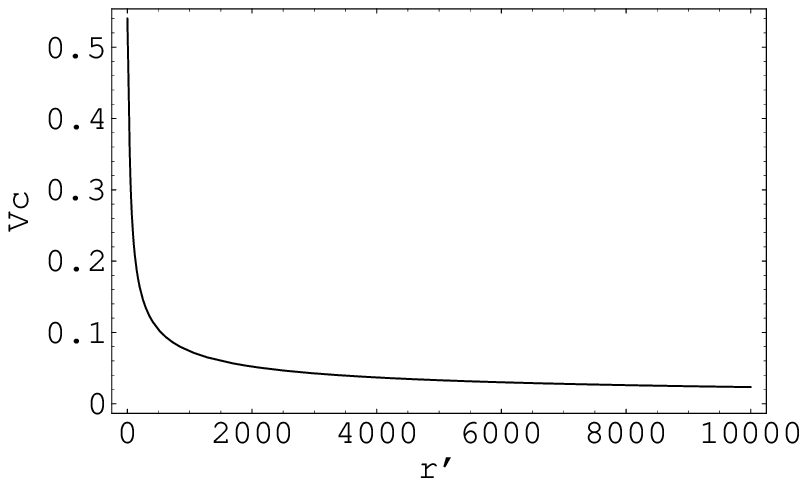}
\end{center} 
\caption{\small {\bf (a)} Disk rotation curves  with
extra dimensional parameters $C_x=0$, $C_y=0$, i.e., usual standard relativistic profile (dotted line, stable),
; $C_x=0.1$, $C_y=0.7$ (dashed line, unstable); $C_x=0.1$, $C_y=0.85$ (gray line, stable); $C_x=0.1$, $C_y=0.9$ (full line, stable). {\bf (b)} Only the stable curves, for various values of the cut parameter $a$; here $a$ is varying from $a=1$ to $a=2.5$, and $C_x = 0.1$ and $C_y=0.85$. The dotted ones are the Newtonian-like (where there are no extra dimensions), and the full ones are the stable curves derived admitting two extra dimensions. In {\bf (c)} we show that the curves derived are asymptotically flat, and in large scale they tend to zero. For all, the solutions used are Schwarzschild for the $4D$ part and Chazy-Curzon for extra dimensions. We take $r'=r/m$.} \label{rot_schw_schw}
\end{figure*}

\begin{figure}
\centering
\includegraphics[width=8cm]{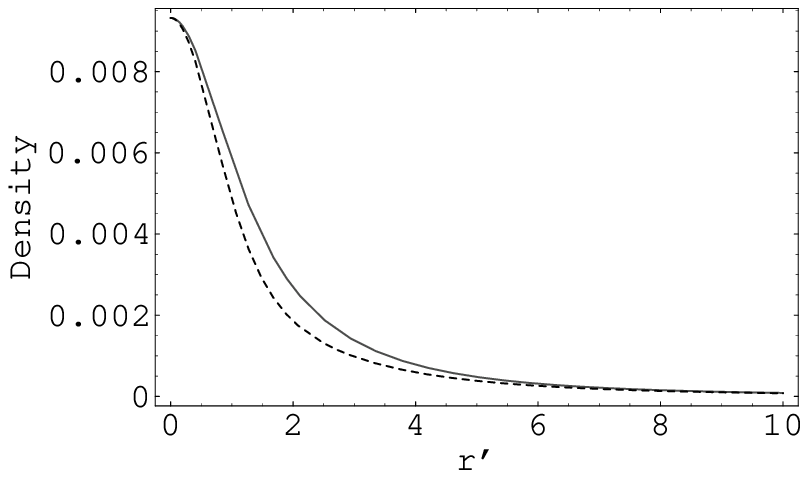}
  \caption{\small Comparison of the density profiles of a disk considering only $4D$ dimensions (full line) and what is derived by the present example that considers $6D$ (dotted line). It is important to point that it is exactly due to this tiny disagreement that extra dimensions do affect rotation curves. Here a stable example with $a=1.5$, for $m=1$ (i.e. $r'=r/m$), $C_x=0.1$ and $C_y=0.85$. From Eqs. (\ref{p1}) and (\ref{f}), the $6D$ density depends on $\phi_{,z}/r'$, which clearly appears due to the influence of extra dimensions. For a large $r'$, $\phi_{,z}/r'$ goes to zero and Eq. (\ref{p1}) recovers the original $4D$ profile. When $r' \rightarrow 0$, $\phi_{,z}/r' \rightarrow \infty$, but $f$ goes faster to infinity, implicating that also  Eq. (\ref{p1}) recovers again the original $4D$ profile for small $r'$.}
\label{fig:comparison}
\end{figure}

The stability study of the previous section is fundamental to discuss what are the exact values to be used for the extradimensional 
parameters $C_x$ and $C_y$. In this sense, it is possible to derive a semi-phenomenological model where we have a very short range of stable values that can be used to compare to rotation curves of galaxies. In Fig. \ref{rot_schw_schw}a we show  the rotation curves for some values of $C_x$ and $C_y$, where we are not pondering yet the stable cases. In \ref{rot_schw_schw}b we present only the stable curves, for various stable values of the cut parameter $a$. As discussed before, the stability is only achieved for $0<C_x<0.4$ and $0<C_y<0.95$ (Rayleigh criterion) or the values from the perturbative method specified in Table \ref{table1} (what restringes even more the range for $C_x$ and confirms the Rayleigh criterion for $C_y$). At the same time those values also prevent superluminal behavior for the particles. When $C_x=C_y=0$  we have the usual $4D$ general relativistic profile that is quite similar to a typical Newtonian one. On the other hand, the ``flattening'' of galactic rotation curves is an effect observed to occur in the galactic area that contains most of the baryonic (visible) matter. Beyond this area it becomes extremely difficult to ascertain the behavior of rotation profiles, so a model that just keeps these rotation profiles flat may be incorrect at larger cosmological scales. In Fig. \ref{rot_schw_schw}c we show that the curves derived are asymptotically flat and in cosmological scales the curves tend to zero.
The density profiles as discussed previously are practically the same as in the $4D$ case and we find that the extra dimensions affect smally the density and the azimuthal and radial pressures. The $4D$ profiles can be seen in Fig. \ref{fig:4dpressures} and in Bi\v c\'ak et al. \cite{bicak}. In Fig. \ref{fig:comparison} we show a comparison between the profiles derived with and without extradimensional imprints, showing that the disagreement is small and the profile derived from the present model is similar to a $4D$ one. Nevertheless, it is exactly due to a such tiny disagreement, induced by the geometry of extra dimensions at large distances, that we have a correct profile for rotation curves. In other words, the presence of extra dimensions is affecting gravity at large distances. Maybe this effect is connected to theories on modified gravity \cite{mond}, and the real nature of such modifications is due to extra dimensions.

\begin{figure*}
\begin{center}
$\begin{array}{c@{\hspace{0.01in}}c} \multicolumn{1}{l}{\mbox{\bf
(a)}} &
    \multicolumn{1}{l}{\mbox{\bf (b)}} \\ [-0.33cm]
\epsfxsize=3.5in \epsffile{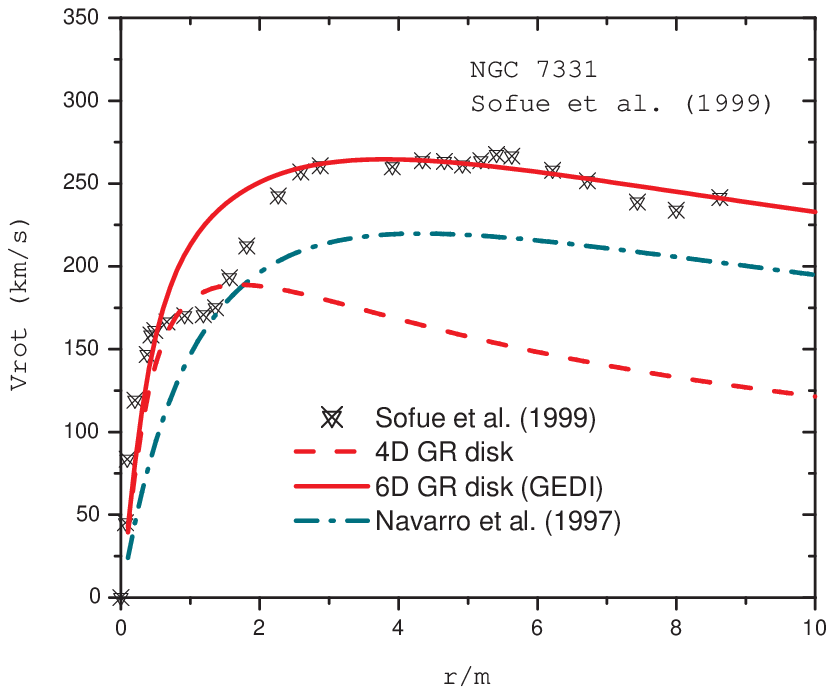} &
    \epsfxsize=3.5in
    \epsffile{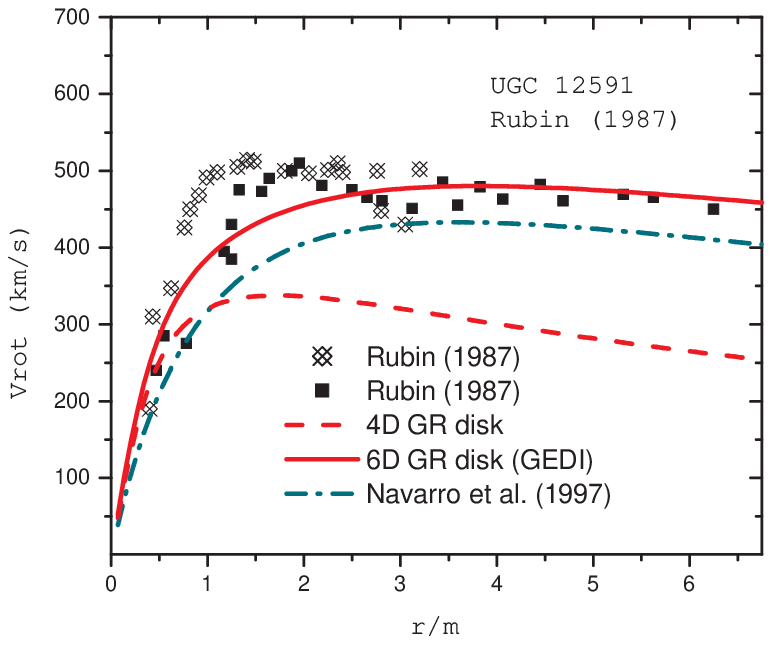} \\ [0.05cm]
\end{array}$
\hspace{0.01in}\mbox{\bf (c)}\\\epsfxsize=3.5in
    \epsffile{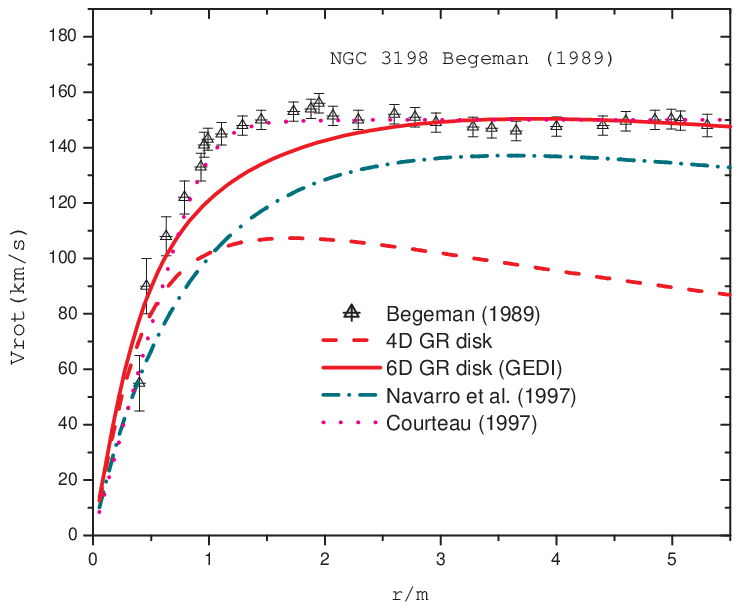}
\end{center}
\caption{\small {\bf (a)} Different models --- our model (a Gravity with Extra Dimensions (GEDI) model) and \cite{navarro}) for the rotation curves of the NGC 7331 spiral galaxy. Observational data from Sofue et al. (1999) \cite{sofue}. {\bf (b)} Different models (our model GEDI and \cite{navarro}) for rotation curves of the high speed rotation spiral galaxy UGC 12591. Observational data from Rubin (1987) \cite{rubin}. {\bf (c)} Different models (our GEDI model and \cite{navarro,courteau}) for the rotation curves of the NGC 3198 spiral galaxy. Observational data from Begeman (1999) \cite{begeman}. The model fits with great precision the region of interest -- the plateau anomaly after $r/m \sim 3$ in {\bf (a)} and after $r/m \sim 2$ in {\bf (b)} and {\bf (c)}. The stable parameters used are $C_x=0.15$ and $C_y=0.88$ for {\bf (a)} and {\bf (b)}, and $C_x=0.2$ and $C_y=0.85$ for {\bf (c)}. For all, the solutions used are Schwarzschild for the $4D$ part and Chazy-Curzon for extra dimensions. The Navarro-Frenk-White \cite{navarro} profile describes the halo curve and when combined with the disk gas curve it phenomenologically fits the rotation curve. The Courteau curve in {\bf (c)} is only a reference of a pure phenomenological fit to rotation curves.} \label{fig:spiral_rotation}
\end{figure*}

Taking into account only the stable curves, in Fig. \ref{fig:spiral_rotation} we compare with some optically observed rotation curves of spiral galaxies. 
Here we are not doing a composition of a halo dark matter velocities plus the disk gas and the velocities of stars. What is happening is that the clean stable general relativistic disk geodesics are simply fitting the region of interest. The not surprising {\it ad hoc} adjustment of $C_x$ and $C_y$ actually could tell nothing about the astrophysical role of the extra dimensions in the model. However the calculation of stable disks brings over with it realizable values for $C_x$ and $C_y$, what makes possible to visualize a minimum representation of a real disk galaxy. Those values produce the full line curves of Fig. \ref{rot_schw_schw}b, fitting with great precision the region of interest (the plateau anomaly after $r' \sim 2$).  We also compare to some phenomenological models used in astrophysics as Navarro, Frenk \& White \cite{navarro} and the Courteau fit \cite{courteau}. The observed data is taken from \cite{sofue,rubin,begeman}, but for an alternative data source see \cite{rotation}, where our model is also successful. Another important statement is that the present model recovers the
Newtonian profile for $r' \rightarrow \infty$ (where the asymptotic function $\nu \rightarrow 0$, and the Newtonian limit is reached).

It is important to remember that this is not a pure phenomenological model, where we would be interested in take a complete fit of observational curve. Here the interest is to achieve a reasonable explanation for the plateau anomaly using some of the stable calculated parameters. In this sense, the model could be considered as a semi-phenomenological model. 
\section{Results for solar system scales}\label{solar}
By simple quantitative arguments, one can easily show that, at solar system scales, the model presents the correct observed results for rotation curves (a Newtonian-like behavior). The argument remains basically on the comparison between distance scales and the mass of the system. Consider, for simplicity, as an example based on observational real quantities, that a spiral galaxy has $M_{gal} \sim 10^{12} M_{\odot} \sim 10^{42}$ kg, with radius $R_{gal} \sim 10$ kpc $\sim 10^{20}$ m. The normalization scale is then $R'_{gal} = R_{gal}/M_{gal} = 10^{-22}$ m/kg. This gives, in Fig. \ref{rot_schw_schw}a, that the $r'$ scale (in natural units) for a such galaxy is around $r' \sim 100$ (where the observed rotation curve is around $r' \sim 10$). Now, doing the same estimative for solar system we have, considering roughly that the solar system mass is $M_{solar} \sim 10 M_{\odot} \sim 10^{31}$ kg and the radius $R_{solar} \sim 1$ pc $\sim 10^{16}$ m, $R'_{solar} = R_{solar}/M_{solar} = 10^{-15}$ m/kg, or $R'_{solar} = 10^{7} R'_{gal}$. Now, the equivalent $r'$ radius for solar system is $r' = 10^{9}$. This last value remains in a absolutely Newtonian-like range for rotation curves (see Fig. \ref{rot_schw_schw}c). Thus the immediate conclusion is that our model is in a correct fashion for both galaxies scales and the solar system scale, using the \emph{same stable parameters for $C_x$ and $C_y$ obtained from stability considerations}. The model, or indeed extra dimensions, is then sensitive to R/M scales.

\begin{figure*}
\centering
\includegraphics[width=10cm]{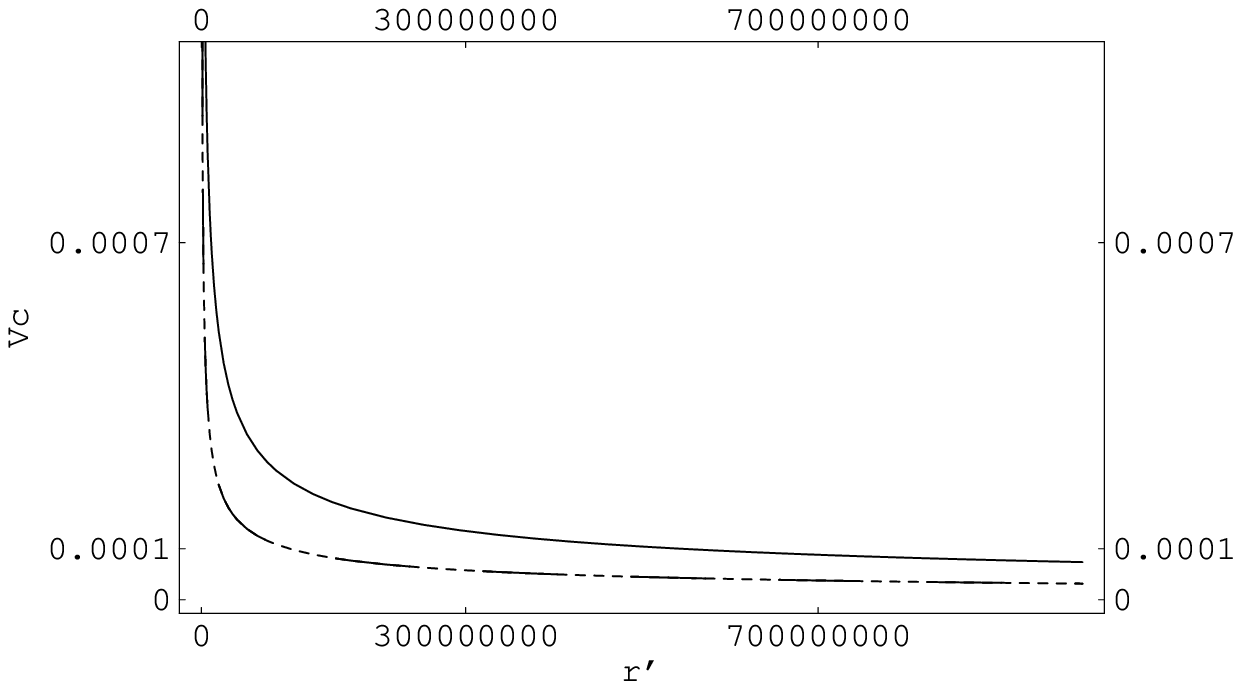}
  \caption{\small Solar system scales. Comparison of rotation curves of a disk considering only $4D$ dimensions, in a Newtonian $\sqrt{1/r'}$ solar system regime (dotted line, which coincides exactly to a $4D$ disk regime --- the dashed line) and what is derived by the present example that considers $6D$ (full line). Here a stable example with $a=1.5$, $r'=r/m$, $C_x=0.1$ and $C_y=0.85$. For a $6D$ disk we have the same profile, but with greater velocities, although the difference is relatively small. The estimative for solar system, considering roughly that the solar system mass is $M_{solar} \sim 10 M_{\odot} \sim 10^{31}$ kg and the radius $R_{solar} \sim 1$ pc $\sim 10^{16}$ m, $R'_{solar} = R_{solar}/M_{solar} = 10^{-15}$ m/kg, or equivalently, in natural units, $r'$ radius for solar system is $r' \sim 10^{9}$.}
\label{fig:solar}
\end{figure*}

In Fig. \ref{fig:solar} we show that all the curves of the model present the same result at solar system scale (the full line) and the dotted line is the Newtonian $\sqrt{1/r'}$ profile (or what is approximately observed at solar system scales). The difference between our model and what is expected for a Newtonian disk is relatively small ($\sim 10^{-5}$).
\section{Cluster of galaxies}\label{sec:lensing}
Here, using the simple example for $6D$, we show that  in the low acceleration regime, a universe endowed with UED predicts gravitational lensing of the correct magnitude to explain the observations of intergalactic lensing. As in nature many elliptical galaxies and galaxy clusters are well modeled as spherically symmetric, we do our calculation for a spherically symmetric system plus two UED. We adopt the metric
\ba \label{spheric}
\mathrm{d}s^2=-e^{\Phi}\mathrm{d}t^2&+&e^\Lambda[\mathrm{d}R^2+R^2(\mathrm{d}\theta^2+\sin^2\theta\mathrm{d}\varphi^2)\nonumber\\
&+&\mathrm{d}x^2 + \mathrm{d}y^2],
\ea where, as before, $x$ and $y$ are the flat extradimensional coordinates. $\Phi$ and $\Lambda$ are functions only of $R$. Consider a light ray which propagates in the equatorial plane of the metric.  The velocity $\dot x^A$ of the ray must satisfy
\be \label{ray}
-e^{\Phi}\, \dot t^2+e^{\Lambda}(\dot{R}^2+R^2\dot\varphi^2+\dot{x}^2+\dot{y}^2)=0.
\ee From the stationarity of the metric follows the conservation law $e^{\Phi} \dot t=E$ where $E$ is a constant characteristic of the ray.  From spherical symmetry it follows that $e^{\Lambda}R^2\dot\varphi=L$ where $L$ is another constant property of the ray. Let us write $\dot R=(\mathrm{d}R/\mathrm{d}\varphi) \dot\varphi $, $\dot x=(\mathrm{d}x/\mathrm{d}\varphi) \dot\varphi$ and $\dot y=(\mathrm{d}y/\mathrm{d}\varphi) \dot\varphi$. Now eliminating $\dot t$ and $\dot\varphi$ from Eq.~(\ref{ray}) in favor of $E$ and $L$, and dividing by $E^2$ yields
\ba
-e^{-\Phi}&+&(b/R)^2
e^{-\Lambda}\{R^{-2}[(\mathrm{d}R/\mathrm{d}\varphi)^2+(\mathrm{d}x/\mathrm{d}\varphi)^2\nonumber\\&+&(\mathrm{d}y/\mathrm{d}\varphi)^2]+1\}=0,
\ea where $b\equiv L/E$. By going to infinity where the metric factors approach unity one sees that $b$ is just the impact parameter of the ray with respect to the matter $4D$ distribution center at $R=0$. Rearranging the last equation we obtain the quadrature
\be
\varphi= [1+R_x^2+R_y^2]^{1/2}\int^R\left[e^{\Lambda-\Phi}\left(\frac{R}{b}\right)^2-1\right]^{-1/2} \frac{\mathrm{d}R}{R},
\ee where $R_x^2=(\mathrm{d}x/\mathrm{d}R)^2$ and $R_y^2=(\mathrm{d}y/\mathrm{d}R)^2$ are the rates where the KK imprints are distributed along the $4D$ galaxy cluster. At this point, where the physical metric exactly flat, this relation would describe a line with $\varphi$ varying from $0$ to $\pi$ as $R$ decreased from infinity to its value $R_{turn}$ at the turning point, and then returned to infinity.  Therefore the deflection of the ray due to gravity is
\begin{widetext}
\be
\Delta\varphi=[1+R_x^2+R_y^2]^{1/2}\left\{2\;\int^\infty_{R_{turn}}
\left[e^{\Lambda-\Phi}\left(\frac{R}{b}\right)^2-1\right]^{-1/2}\frac{\mathrm{d}R}{R}-\pi\right\}.
\ee 

To shed some light in this last integral, one can benefit of the weakness of extragalactic fields which grant that $\Lambda$ and $\Phi$ are all small compared to unity.  As consequence the above result is closely approximated by        
\be
\Delta\varphi=[1+R_x^2+R_y^2]^{1/2}\left\{-4\;\frac{\partial}{\partial\alpha}\int^\infty_{R_{turn}}
\left[(1+\Lambda-\Phi)\left(\frac{R}{b}\right)^2-\alpha\right]^{1/2} \frac{\mathrm{d}R}{R}\Big|_{\alpha=1}-\pi\right\}.
\ee 
\noindent The rewriting in terms of an $\alpha$ derivative allows us to Taylor expand the radical in the small quantity $\Lambda-\Phi$ without incurring a divergence of the integral at its lower limit.  The zeroth order of the expansion yields a well known integral which cancels the $\pi$.  Thus, to first order in small quantities 
\be
\Delta\varphi=-{2\;[1+R_x^2+R_y^2]^{1/2}\over
b}{\partial\over\partial\alpha}\int^\infty_{b\surd\alpha}
{(\Lambda-\Phi)R \mathrm{d}R\over(R^2-\alpha
b^2)^{1/2}}\Big|_{\alpha=1}.
\ee And integrating by parts:
\be \label{firstint}
\Delta\varphi=-{2\;[1+R_x^2+R_y^2]^{1/2}\over
b}{\partial\over\partial\alpha}\Big[\lim_{R\rightarrow
\infty} (\Lambda-\Phi)(R^2-\alpha
b^2)^{1/2}-\int_{b\surd
\alpha}^\infty (\Lambda_{,R}-\Phi_{,R})(R^2-\alpha b^2)^{1/2}
\mathrm{d}R\Big]\Big|_{\alpha=1}
\ee 
\end{widetext}
Since $\Phi$ and $\Lambda$ decrease asymptotically as $R^{-1}$, the integrated term, being $\alpha$ independent, contributes nothing.  Carrying out the $\alpha$ derivative, and introducing the usual Cartesian $u$ coordinate along the initial ray by $u\equiv \pm (R^2-b^2)^{1/2}$, we have
\be \label{lensing}
\Delta\varphi={b\;[1+R_x^2+R_y^2]^{1/2}\over 2}\int_{-\infty}^\infty
{\Lambda_{,R} - \Phi_{,R}\over R}\, \mathrm{d}u.
\ee  A factor $1/2$ appears  because we have included the integral in Eq.~(\ref{firstint}) twice, once with $R$ decreasing to, and once with $R$ increasing from $b$.  The integral is now performed over an infinite straight line following the original ray. 

\begin{figure}
\centering
\includegraphics[width=8cm]{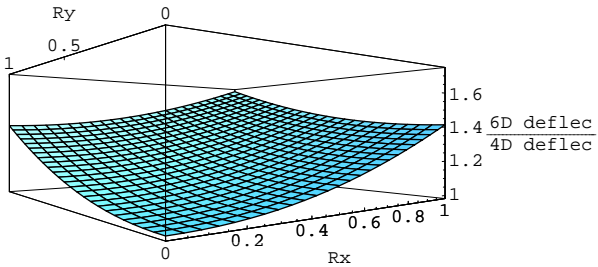}
  \caption{\small The new ray deflection using the above calculated model, for a range of projected rates $R_x$ and $R_y$, is bigger than the deflection produced by a ray passing through a cluster calculated only with $4D$ general relativity (GR). The presence of extra dimensions acts exactly as a dark halo of cold dark matter.}
\label{fig:lensing1}
\end{figure}

Fig. \ref{fig:lensing1} shows that the new ray deflection using the above calculated model, for a range of projected rates $R_x$ and $R_y$, is bigger than the deflection produced by a ray passing through a cluster calculated only with $4D$ general relativity (GR).
The difference between a $4D$ GR with dark matter and the above calculation
(where the geometry of the problem substitutes the concept of DM) is that in a conventional dark matter scenario one would compute $\Phi$ and $\Lambda$ from Einstein equations including dark matter as source, whereas in the present model one has an aditional term and computes $\Lambda$ and $\Phi$ on the basis of the visible matter alone.

\begin{figure}
\centering
\includegraphics[width=9cm]{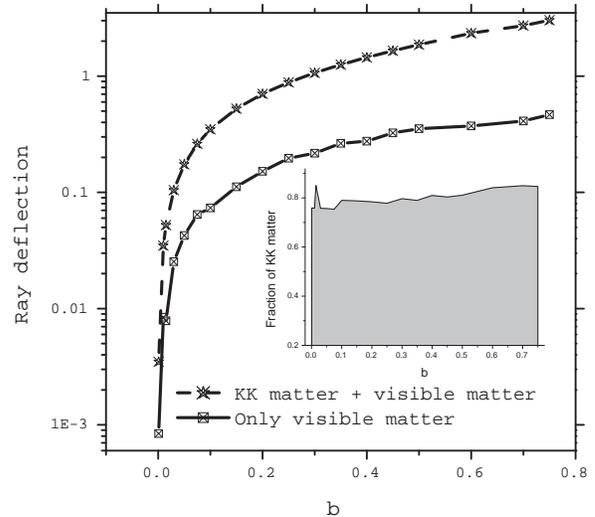}
  \caption{\small Numerical integration of the gravitational lensing deflection produced by a cluster living in a universe endowed with UED. Here we use a equivalence between the gravitational field induced by extra dimensions and a fictitious matter --- ``KK matter'' --- to show that the contribution of the ``extra'' field induced by extra dimensions acts as dark matter for a galactic cluster. In the example above, the fictitious ``KK matter'' constitutes $(79.5\pm3.3)\%$ of the total matter in a spherically symmetric galactic cluster. We used $R_y=0.9$ and $R_x=0.4$, stable values according to the disk problem.}\label{fig:lensing2}
\end{figure}

Solving numerically the Einstein equations of the spherical metric and by a numerical integration of Eq. (\ref{lensing}), it is possible to obtain a comparison between a galactic cluster with and without extra dimensions. In Fig. \ref{fig:lensing2} we demonstrate this, using the same values for $C_x$ and $C_y$ calculated in Sec. \ref{sec:stability}. In our model, for the cluster scales, extra dimensions geometry deflect more light as awaited. Thus the system is equivalent to a DM system where $(79.5\pm3.3)\%$ of the total matter in a spherically symmetric galactic cluster is ``dark''. This is very similar to what happens accordingly to observations. It is verified by lensing observations \cite{oort} that cluster of galaxies are composed of three main components: $\sim 5\%$ in mass is the optically luminous baryonic matter in hundred of bright galaxies; $\sim 10$--$15\%$ is in the form of a bright X-ray inter-cluster gas; and the remaining $\sim 80$--$85\%$ is some sort of non-baryonic ``missing mass''. The explanation of our results is that such non-baryonic ``missing mass'' can be explained by the geometrical presence of Universal Extra Dimensions. One can picture the problem as the geometry provides an ``equivalent'' residual gravitational mass. The ``extra'' gravitational field is induced only by extra dimensions and observationally a such induction is seen as a missing mass.

Actually, using simple quantitative arguments as in Sec. \ref{solar}, one can see that a typical cluster of mass $M_{\mathrm{cluster}} \sim 10^{15} M_{\odot} \sim 10^{45}$ kg and radius $R_{\mathrm{cluster}} \sim 10$ Mpc $\sim 10^{23}$ m, has R/M scales of order $R' \sim 10^{-22}$ m/kg, similar to the values calculated for galaxies. In consequence, it is reasonable to expect a strange modified behavior for cluster similar as those for galaxies. The final conclusion is that a multidimensional universe is sensitive to R/M scales.

\section{Concluding remarks}\label{sec:discussion}
At the present paper, the possibility to construct a galactic disk embedded in a multidimensional space-time was investigated. Four principal results was achieved: i) The density profiles of the disk have a $4D$ similar shape, but present a very soft difference, indicating that extra dimensions are modifying the gravitational field at galaxy scales ($r' \sim 1-100$). ii) The rotation curves of stable disks fit the profile of observed galaxies. iii) The rotation curves for solar system scales are Newtonian-like, as observed astronomically. iv) A cluster of galaxies present a larger deflection for light rays than what is calculated for a $4D$ cluster. 

To achieve such results, after demonstrate that vacuum Einstein equations can be written as a set of Laplace's equations for any number $D$ of dimensions, we showed that if $D$ is even, then the sum of extradimensional pressures cancel and the total pressure depends only of $4D$ components. Thus, the density profiles have practically the same shape as in the $4D$ case and we find that the extra dimensions affect little the azimuthal and radial pressures. As the simplest example, we calculated the case for a six  dimensional space-time. Two  constants of motion from projection of extradimensional particle velocities are the free parameters of the model. We had prevented the {\it ad hoc} adjustment of such parameters with observed rotation curves, preferring to investigate values where the disk becomes stable. The stability is achieved when the disk is Newtonian (where such parameters are null) or for a tiny range of values that makes the circular geodesics fit with great precision the rotation curves of many  spiral galaxies. The stability calculation was done using both the Rayleigh criterion and a perturbative approach. We compared such results to well succeeded astrophysical dark matter profiles and demonstrated that our predictions give the same gravitational lensing as does a dynamically successful dark halo model. We also introduced quantitative arguments to show that at solar system scales, the model presents the correct behavior for rotation curves (a Newtonian-like profile).

The method to construct the disk is not new. Actually we used well known general relativistic methods to construct thin disks, including, however, as a novelty, extra dimensions in the geometry of the system. Though the presented results introduce important constraints to Kaluza-Klein theories. If it is true, maybe it would mean that the world we perceive is only a fraction of a greater invisible reality and maybe the $4D$ matter is a kind of modulation demanded both by the curvature of $4D$ space-time and the extradimensional geometry.

Two main questions could be formulated about the results of the present paper: i) Do such results bring an alternative to the dark matter conjecture (showing that the strenght of gravitation may be modified by extra dimensions, which would have similar consequences)? ii) Or do they rather suggest a particular form of dark matter (consisting in Kaluza-Klein particles that may result from extra dimensions)? A first observation on this aspect is that modifications in rotation curves come from the possibility that extra dimensions are endowed with dynamical test particles, i.e., the extra dimensions must contain the presence of fields (as similar to Universal Extra Dimensions models). Eqs. (\ref{cxcy}) and (\ref{ddr}) show that the projection of extra dimensional rates is the determinant factor to obtain a deviation from a Newtonian profile. Becomes clear, from Secs. \ref{solar} and \ref{sec:lensing}, that the effect of extra dimensions is to modify gravity according to the astronomical scale R/M (distance per mass), thus altering the rotation curves and lensing, for instance, in a more rigid framework than that of CDM models. In this sense, our model is similar to e.g. MOND or TeVeS \cite{mond}, but proposes and demonstrates that modifications of gravity comes naturally from the geometry of a multidimensional universe and that a multidimensional space-time is sensitive to R/M (distance per mass) scales.

\acknowledgments
C.H. C.-A. thanks IUPAP and CAPES and P.S.L
thanks CNPq and FAPESP for partial financial support. The authors are also very grateful to PRD referees for enlightening views and for suggestions given in order to improve the quality of this paper.\\

\section*{Appendix: An alternative for the extradimensional modified gravity interpretation}
Alternatively, we can also present another interpretation for the nature of the results. One can assume that actually the model is constraining a Kaluza-Klein dark matter particle (provenient from UED imprints). See detailed discussions about a such topic in \cite{kkdm}. And maybe both interpretations (the pure geometrical approach and the induced KK matter), are physically equivalent (see for instance a idea proposed by Clifford in \cite{clifford}, where not only could one describe fields through geometry, but also the particles that interact by means of those fields). About the particle aspect, although the main purpose of the present work is to construct a top-down model and not to shed light on a fundamental theory we can illustrate considering the possibility that our model could constrain a Kaluza-Klein dark matter particle to be tested at Large Hadron Collider (LHC) in next years. As an example, consider for simplicity that Kaluza-Klein dark matter particle in Universal Extra Dimensions could be considered as a fundamental particle, where is included also the possibility that a such particle could decay in others. Following \cite{acd}, one can derive a Lagrangian from the compactification of the extra dimensions and connect such compactification to the cross section of the particle. Using the generic notation $x^\alpha, \; \alpha = 0, 1, ..., 3 +\delta$ for the coordinates of the $(4+\delta)$-dimensional space-time (different from the usual non-compact space-time coordinates, $x^\mu, \, \mu = 0,1,2,3$), and the coordinates of the extra dimensions, $y^a, \, a = 1, ..., \delta$, the 4-dimensional Lagrangian can be obtained by dimensional reduction from the $(4+\delta)$-dimensional theory, 
\begin{widetext}
\ba {\cal L}(x^\mu) & = & \int d^{\delta} y \left\{ -
\sum_{i=1}^3 \frac{1}{2 \hat{g}_i^2} {\rm
Tr}\left[F_i^{\alpha\beta}(x^\mu, y^a)
{F_i}_{\,\alpha\beta}(x^\mu, y^a)\right]  + {\cal L}_{\rm
Higgs}(x^\mu,y^a) \right. \nonumber \\ [2mm] &+& \left. \!\!\!\! i
\left(\overline{\CQ}, \overline{\CU}, \overline{\CD}\right)
(x^\mu, y^a) \left(\Gamma^\mu D_\mu + \Gamma^{3+a} D_{3+a} \right)
\left(\CQ, \CU, \CD \right)^\top (x^\mu, y^a) \right.
\nonumber \\ [2mm] &+& \left. \!\!\!\!\!
\left[\overline{\CQ}(x^\mu,y^a)\left(\hat{\lambda}_\CU
\CU(x^\mu,y^a) i\sigma_2 H^*(x^\mu,y^a) +
 \hat{\lambda}_\CD \CD(x^\mu, y^a) H(x^\mu, y^a)\right) +
{\rm h.c.} \right] \right\} ~. \label{lagrangian} \ea  
\end{widetext}
Here $F^{\alpha\beta}_i$ are the $(4+\delta)$-dimensional gauge field strengths associated with the $SU(3)_C \times SU(2)_W \times U(1)_Y$ group, while $D_\mu = \partial/\partial x^\mu - \CA_\mu$ and $D_{3+a} = \partial/\partial y^a - \CA_{3+a}$ are the covariant derivatives, with $\CA_\alpha = -i \sum_{i=1}^3\hat{g}_i {\CA_\alpha^{r}}_i T^{r}_i$ being the $(4+\delta)$-dimensional gauge fields. The piece ${\cal L}_{\rm Higgs}$ of the $(4+\delta)$-dimensional Lagrangian contains the kinetic term for the $(4+\delta)$-dimensional Higgs doublet $H$, and the Higgs potential. The $(4+\delta)$-dimensional gauge couplings $\hat{g}_i$, and the  Yukawa couplings collected in the $3\times 3$ matrices $\hat{\lambda}_{\CU,\CD}$, have dimension (mass)$^{-\delta/2}$. The fields $\CQ, \CU$ and $\CD$ describe the $(4+\delta)$-dimensional fermions whose zero-modes are given by the 4-dimensional standard model quarks. A summation over a generational index is implicit in Eq.~(\ref{lagrangian}). For example, the 4-dimensional, third generation quarks may be written as $\CQ^{(0)}_3 \equiv (t, b)_L, \, \CU^{(0)}_3 \equiv t_R, \,\CD^{(0)}_3 \equiv b_R$. The kinetic and Yukawa terms for the weak-doublet and -singlet leptons, $\CL$ and $\CE$, are not shown for brevity. The gamma matrices in $(4+\delta)$ dimensions, $\Gamma^\alpha$, are anti-commuting $2^{k + 2}\times 2^{k + 2}$ matrices, where $k$ is an integer such that $\delta = 2k$ or $\delta = 2k+1$. Chiral fermions exist only when $\delta$ is even, and correspond to the eigenvalues $\pm 1$ of $\Gamma^{4+\delta}$. The space-time is described as $M^4 \times T^\delta$, where the extra dimensions can be compactified in a $T^\delta$ tori. The $6D$ example of Section \ref{sec:6D} has a tori of radii $r_x= e^{\nu/2}$ and $r_y= e^{-\nu/2}$. Asymptotically the compactification tori radius is constant and it is obtained by $r_c= \sqrt{r_x^2+r_y^2}$, and the periodic dimensions $x^k \cong x^k + 2 \pi r_c$, $6-\delta \leq k \leq 5$. The compactification is obtained by imposing the identification of two pairs of adjacent sides of the torus retangle. In a pure UED model the compactication scale arises simply from the need to produce the right amount of DM. Here, a modified UED model, the compactification scale arises from the dynamics of galaxies and from clusters lensing. A detailed investigation about the compactification of the model will be done in future works, although our preliminary results indeed strength the possibility that our model could constrain, as one of the physical interpretations, a Kaluza-Klein dark matter particle to be tested at Large Hadron Collider (LHC), where the proposed energy range is 1--14 TeV, and the main signals of KK DM will be several $t\bar{t}$ resonances.

\end{document}